\documentclass[12pt]{article}
\usepackage{epsf,rotate,amssymb}
\topmargin by -\headsep \textheight 8.9in \oddsidemargin 0pt
\evensidemargin \oddsidemargin \marginparwidth 0.5in \textwidth
6.5in 

\newcommand{\rb}[1]{\raisebox{1.8ex}[0pt]{#1}}

\newcommand{\e}{\epsilon}
\newcommand{\bi}{\begin{itemize}}
\newcommand{\ei}{\end{itemize}}

\newcommand{\dlmax}{d_{l\mbox{max}}}
\newcommand{\onen}{\frac{1}{n}}

\newcommand{\AnL}{\Acal_{n\mbox{{\rm \tiny L}}}}
\newcommand{\AoneL}{\Acal_{1\mbox{{\rm \tiny L}}}}
\newcommand{\AL}{\Acal_{\mbox{{\rm \tiny L}}}}

\newcommand{\AnT}{\Acal_{n\mbox{{\rm \tiny T}}}}

\newcommand{\AT}{\Acal_{\mbox{{\rm \tiny T}}}}

\newcommand{\AnTP}{\Acal_{n\mbox{{\rm \tiny TP}}}}

\newcommand{\ATP}{\Acal_{\mbox{{\rm \tiny TP}}}}

\newcommand{\AcalSol}{\overline{\Acal^*}}

\newcommand{\AnD}{\Acal_{n\mbox{{\rm \tiny D}}}}

\newcommand{\AD}{\Acal_{\mbox{{\rm \tiny D}}}}

\newcommand{\AnDP}{\Acal_{n\mbox{{\rm \tiny DP}}}}

\newcommand{\ADP}{\Acal_{\mbox{{\rm \tiny DP}}}}

\newcommand{\AnDPC}{\Acal_{n\mbox{{\rm \tiny DPC}}}}

\newcommand{\ADPC}{\Acal_{\mbox{{\rm \tiny DPC}}}}

\newcommand{\AnLPC}{\Acal_{n\mbox{{\rm \tiny LPC}}}}

\newcommand{\ALPC}{\Acal_{\mbox{{\rm \tiny LPC}}}}

\newcommand{\AnTPC}{\Acal_{n\mbox{{\rm \tiny TPC}}}}

\newcommand{\ATPC}{\Acal_{\mbox{{\rm \tiny TPC}}}}

\newcommand{\AnDC}{\Acal_{n\mbox{{\rm \tiny DC}}}}

\newcommand{\AnTC}{\Acal_{n\mbox{{\rm \tiny TC}}}}

\newcommand{\AnE}{\Acal_{n\mbox{{\rm \tiny E}}}}

\newcommand{\Ae}{\Acal_{\mbox{{\rm \tiny E}}}}

\newcommand{\AnLC}{\Acal_{n\mbox{{\rm \tiny LC}}}}
\newcommand{\AoneLC}{\Acal_{1\mbox{{\rm \tiny LC}}}}
\newcommand{\ALC}{\Acal_{\mbox{{\rm \tiny LC}}}}

\newcommand{\ALCol}{\overline{\Acal^*_{\mbox{{\rm \tiny LC}}}}}

\newcommand{\AnLP}{\Acal_{n\mbox{{\rm \tiny LP}}}}

\newcommand{\ALP}{\Acal_{\mbox{{\rm \tiny LP}}}}

\newcommand{\Tenn}{{\cal T}_{\epsilon'}^{(n')}}

\newcommand{\xol}{\mbox{$\overline{x}$}}
\newcommand{\Wol}{\mbox{$\overline{W}$}}

\newcommand{\Xcalol}{\overline{\cal X}}
\newcommand{\Ycalol}{\overline{\cal Y}}
\newcommand{\Ucalol}{\overline{\cal U}}
\newcommand{\Wcalol}{\overline{\cal W}}
\newcommand{\Xcal}{{\cal X}}
\newcommand{\Ecal}{{\cal E}}
\newcommand{\Acal}{{\cal A}}
\newcommand{\Vcal}{{\cal V}}

\newcommand{\Bcal}{{\cal B}}
\newcommand{\Zcal}{{\cal Z}}
\newcommand{\Wcal}{{\cal W}}
\newcommand{\Ycal}{{\cal Y}}
\newcommand{\Ucal}{{\cal U}}
\newcommand{\Zcalol}{\overline{\cal Z}}

\newcommand{\Ihat}{\hat{I}}
\newcommand{\Itil}{\tilde{I}}

\newcommand{\Zhat}{\mbox{$\hat{Z}$}}

\newcommand{\xolhat}{\widehat{\overline{x}}}
\newcommand{\Xolhat}{{\widehat{\overline{X}}}}
\newcommand{\Zolhat}{{\widehat{\overline{Z}}}}

\newcommand{\Yolhat}{{\widehat{\overline{Y}}}}
\newcommand{\Zol}{{{\overline{Z}}}}
\newcommand{\Rol}{{{\overline{R}}}}
\newcommand{\Dol}{{{\overline{D}}}}

\newcommand{\wol}{\mbox{$\overline{w}$}}

\newcommand{\Xol}{\mbox{$\overline{X}$}}

\newcommand{\Xhat}{\mbox{$\hat{X}$}}

\newcommand{\Yol}{\mbox{$\overline{Y}$}}

\newcommand{\yol}{\overline{{y}}}

\newcommand{\Scal}{{\cal S}}

\newcommand{\fol}{\mbox{$\overline{f}$}}

\newcommand{\Rbb}{\mathbb{R}}

\newcommand{\E}{\mbox{E}}
\newcommand{\be}{\begin{equation}}
\newcommand{\ee}{\end{equation}}
\newcommand{\bea}{\begin{eqnarray}}
\newcommand{\eea}{\end{eqnarray}}
\newcommand{\beann}{\begin{eqnarray*}}
\newcommand{\eeann}{\end{eqnarray*}}

\newtheorem{theorem}{Theorem}[section]
\newtheorem{corollary}[theorem]{Corollary}
\newtheorem{lemma}[theorem]{Lemma}

\renewcommand{\theequation}{\arabic{section}.\arabic{equation}}

\newcommand{\Section}[1]{\section{#1}
\setcounter{equation}{0}
\setcounter{figure}{0}
\setcounter{table}{0}}

\title{Unified Theory of Source Coding:\\ 
Part II -- Multiterminal Problems}

\author{Soumya Jana\\
        University of Illinois at Urbana-Champaign \\
        Email: {\tt \{jana\}@uiuc.edu}}
\date{}

\begin{document}

\maketitle

\thispagestyle{plain}
\pagestyle{plain}

\baselineskip=1.25\normalbaselineskip
\renewcommand{\baselinestretch}{1.4}

\begin{abstract}
In the first paper of this two part communication, we solved in a unified framework a variety of two terminal source coding problems 
with noncooperative encoders, thereby consolidating works of Shannon, Slepian-Wolf, Wyner, Ahlswede-K\"{o}rner, Wyner-Ziv, Berger {\em et al.} and Berger-Yeung.
To achieve such unification we made use of a fundamental principle that dissociates bulk of the analysis from the distortion criterion at hand (if any) and extends the typicality arguments of Shannon and Wyner-Ziv. In this second paper, we generalize the fundamental principle for any number of sources and  on its basis exhaustively solve all multiterminal source coding problems with noncooperative encoders and one decoder. The distortion criteria, when applicable, are required to apply to single letters and be bounded. Our analysis includes cases where side information is, respectively, partially available, completely available and altogether unavailable at the decoder.
As seen in our first paper, the achievable regions permit infinite order information-theoretic descriptions. We also show that the entropy-constrained multiterminal estimation problem can be solved as a special case of our theory.
\end{abstract}

\renewcommand{\baselinestretch}{1.5}

\newpage

\Section{Introduction}
\label{sec:intro}
In the first paper of this two part communication \cite{PartI}, we solved in a unified framework a variety of two terminal source coding problems 
with noncooperative encoders, consolidating works of Shannon \cite{ShanLL,Shannon}, Slepian-Wolf \cite{SW}, Wyner \cite{Wyner}, Ahlswede-K\"{o}rner \cite{AhlKor}, Wyner-Ziv \cite{WZ}, Berger {\em et al.} \cite{Upper} and Berger-Yeung \cite{BY}. In particular, we derived a fundamental source coding principle extending the typicality arguments of Shannon \cite{Shannon} and Wyner-Ziv \cite{WZ}, and, using this principle, showed inner bound properties on the achievable regions. We also showed the outer bound properties using interposed lossless coding (as seen in \cite{WZ}) and Fano's inequality \cite{Cover}. 
In this second paper, we extend our framework to multiterminal source coding with noncooperative encoders. Specifically, we exhaustively enumerate twelve problems in three categories where encoded sources are decoded 1) losslessly, and 2) under distortion criteria, respectively, and  3) a subset of the encoded sources are decoded losslessly whereas the rest are decoded under distortion criteria. In each category, one of the following four subcases arises.
At the decoder either 1) side information is unavailable, or 2) side information is available at a certain rate (partially), or 3) side information is available completely, or 4) part of the side information is available partially whereas part is available completely. We shall see that the eleven remaining problems are special cases of the problem where only a subset of the sources are losslessly decoded whereas the rest are decoded under distortion criteria with part of side information available partially and part available
 completely. This problem sans side information has Berger-Yeung problem \cite{BY} as its two terminal specialization. We solve the general multiterminal version of the above problem using our usual methodology. Specifically, we give an infinite order description of the achievable region. The inner bound is shown using a multiterminal extension of our two terminal fundamental principle that dissociates bulk of the analysis from distortion criteria and extends typicality arguments of Shannon \cite{Shannon} and Wyner-Ziv \cite{WZ}. 
The outer bound is shown using interposed multiterminal lossless coding and Fano's inequality extending our two terminal argument \cite{PartI}. Finally, we shall demonstrate that the scope of our theory extends beyond the traditional source coding. In particular, we shall solve the entropy-constrained estimation problem in a multiterminal setting as a special case of our theory.
We organize our analysis as follows: We pose the twelve multiterminal source coding problems in Sec.~\ref{sec:prob} and present their solutions in Sec.~\ref{sec:results}. In Sec.~\ref{sec:funda}, we state and prove the general multiterminal version of the fundamental principle of source coding. The proof of our general source coding theorem is given in Sec.~\ref{sec:proofM}. We apply our theory to multiterminal entropy-constrained estimation in Sec.~\ref{sec:decision}. Finally, Sec.~\ref{sec:discuss} concludes the paper.

\Section{Multiterminal Problems} 
\label{sec:prob}
We begin with an exhaustive enumeration of multiterminal source coding problems where individual encoders do not cooperate. In the process, we bring out the similarity, the dissimilarity and the interdependency among such problems. We also identify the problems, which have already been solved completely, which have been solved in special cases and for which certain bounds have been found. Subsequently, we shall solve the unsolved problems in their most general setting. For the sake of convenience, we pose distributed source coding problems in a phased manner:
Basic source coding (without side information) in Sec. \ref{sec:source}, source coding with partial side information in Sec. \ref{sec:part} and source coding with complete side information in Sec. \ref{sec:comp}. First we need some notation and the concept of strong typicality.

\subsection{Notation}
\label{sec:notation}
Throughout this paper we denote random variables by uppercase letters such as $X$, $Y$, $Z$, and their alphabets by corresponding script letters ${\cal X}$, ${\cal Y}$, ${\cal Z}$. All alphabets are finite unless otherwise stated. 
By $H(X)$ and $I(X;Y)$, denote 
entropy of $X$ and mutual information between $X$ and $Y$, respectively. 
Further, 
%by $\zerobf_K$ and $\onebf_K$, respectively, denote the $K$-vectors of $0$'s and $1$'s, and 
by $I_K$,
denote the set $\{1,2,...,K\}$. We adopt the convention $I_0= \{\}$.
Also, denote $j+I = \{j+m:m\in I\}$. 
Clearly, $M + I_K = I_{M+K}\setminus I_M$.
Denote the $k$-th element of a sequence by $x(k)$, the corresponding sequence by $\{x(k)\}$ and the collection of all elements indexed by $k_1$ through $k_2$ by $x(k_1; k_2)$. Also write $x^n=x(1;n)$ and $x^n(k) = x(n(k-1)+1;nk)$. 
%and $x^n(k_1;k_2) = x(n (k_1-1) +1; n k_2)$.
Denote vector (collection) $(X_1,X_2,...,X_M)$ of random variables
by $\Xol$, and the corresponding alphabet by $\Xcalol = \Xcal_1\times \Xcal_2\times ... \times \Xcal_M$. Moreover,
denote by $\fol$, the vector of mappings $f_m:\Xcal_m \rightarrow \Zcal_m$, $m\in I_M$. Here the fact that $\fol$ has component functions with distinct 
domains and ranges will sometimes be indicated by 
the symbol `$\fol: \Xcalol \rightarrowtail \Zcalol$'.
For any $I\subseteq I_M$, denote by $\Xol_I$ the vector of $\{X_m: m\in I\}$ and by $\Xcalol_I= \prod_{m\in I}\Xcal_m$ denote the corresponding alphabet. Also,
denote by $\fol_I$ the vector of mappings $\{f_m\}_{m\in I}$. Further, denote $R_I=\sum_{m\in I} R_m$ (note the contrast with $\Rol_I$, the vector of $\{R_m: m\in I\}$). In addition, denote the closure of set $\Acal$ by $\overline{\Acal}$.
Finally, define the 
$\epsilon$--strongly ($\epsilon>0$) typical set
of $X\sim p(x)$ by \cite{Cover}
\be
\label{eq:typical}
{\cal T}_\epsilon^{(n)}(X) = 
\left\{x^n\in \Xcal^n:
\left|\frac{1}{n} N(x|x^n) - p(x)\right|<\frac{\epsilon}{|\Xcal|} ~\mbox{for all}~x\in \Xcal \right\}, 
\ee
where $N(x|x^n)$ denotes the number of occurrences of $x$ in the sequence $x^n$. In this paper, we consider only strong typicality which will henceforth be mentioned simply as typicality.
Consequently, we have, for sufficiently large $n$ (due to strong law of large numbers), 
\be
\label{eq:TPe}
\Pr\{X^n \notin {\cal T}_\epsilon^{(n)}(X)\} \le \epsilon,
\ee 
where $\{X(k)\}$ are drawn i.i.d. $\sim p(x)$. Also if $x^n \in {\cal T}_\epsilon^{(n)}(X)$, then we call $x^n$ a typical sequence. In an analogous manner, the jointly typical set of a collection of random variables $\Xol = (X_1,X_2,...,X_M)$
is defined by (\ref{eq:typical}) with $X$, $x$ and $\Xcal$ replaced by $\Xol$, $\xol$ and $\Xcalol=\Xcal_1\times\Xcal_2\times...\times\Xcal_M$, respectively.

\subsection{Basic Source Coding}
\label{sec:source}

Consider vector of $M$ random variables 
$\Xol = (X_1,X_2,...,X_M)\sim p(\xol)$, components of which are separately encoded and jointly decoded.
Specifically, draw $\{\Xol(k)\}$ i.i.d. $\sim p(\xol)$, encode $\Xol$ using $M$ encoder mappings
\be
\label{eq:fm}
f_{m}:\Xcal_m^n \rightarrow \Zcal_m, \quad m\in I_M
\ee
(i.e., $\fol:\Xcalol^{n} \rightarrowtail \Zcalol$)
for some alphabet $\Zcalol$
and decode using decoder mapping
\be
\label{eq:g}
g: \Zcalol \rightarrow \Xcalol^n.
\ee
We call 
\be
\label{eq:rec}
\Xolhat^n =g (\fol(\Xol^n))
\ee
the estimate  or reconstruction of $\Xol^n$.
Further, a rate $M$-vector $\Rol$ is said to be achievable if for any $\e>0$, there exists (for $n$ sufficiently large) mapping pair $(\fol,g)$  such that 
\be
\label{eq:AR}
\frac{1}{n} \log |\Zcal_m | \le R_m +\e,\quad m\in I_M
\ee
and appropriate error or distortion criteria based on $(\Xol^n,\Xolhat^n)$ are met also within an accuracy of $\e$. 

Depending on such criteria,
we enumerate three sub-problems (and 
assign each a tag, e.g., `L') in the following.

\begin{enumerate}
\item {\bf Lossless Coding (`L'):} $\Xol$ is losslessly decoded (in the sense of Shannon). Specifically, a rate vector $\Rol$ is said to be achievable if for any $\e>0$, (\ref{eq:AR}) holds alongside
\be
\label{eq:ALL}
\Pr\{\Xol^n\ne \Xolhat^n\}\le \e.
\ee
Denote by $\AL$ the set of achievable $\Rol$. This problem has been solved by Shannon \cite{ShanLL} for $M=1$ and by Slepian and Wolf \cite{SW} for general $M$.

\item {\bf Coding under Distortion Criteria (`D'):} $\Xol$ is decoded  
under $L$ bounded distortion criteria $d_l:\Xcalol^{2} \rightarrow [0,d_{l\max}]$, $l\in I_L$.
The achievable set $\AD$ is defined by the set of pairs $(\Rol,\Dol)$ ($\Dol$ being an $L$-vector) such that, 
for any $\e>0$, (\ref{eq:AR}) holds alongside
\be
\label{eq:ALC}
\frac{1}{n} \E d_{ln}(\Xol^{n}, \Xolhat^{n}) \le  D_l + \e,
\quad l\in I_L
\ee
where
$$
d_{ln}(\xol^n,\xolhat^n) = \sum_{k=1}^n d_l(\xol_k,\xolhat_k)
$$
(of course, $d_{ln}:\Xcalol^{2n}\rightarrow[0,n d_{l\max}]$).
 The special case, where $M=1$ and $L=1$, was solved by Shannon \cite{Shannon}. Also, the case, where $M=2$ and $L=1$, was solved in our first paper \cite{PartI} of this series.  

\item {\bf Lossless Coding in a Subset (`T'):} A subset $\Xol_{J}$, $J\subseteq I_M$, of sources $\Xol$, is losslessly decoded and the complementary subset $\Xol_{J^c}$ is decoded  
under $L$ bounded distortion criteria $d_l:\Xcalol_{J^c}^{2} \rightarrow [0,d_{l\max}]$, $l\in I_L$. 
The achievable set $\AT$ is defined by the set of $(\Rol,\Dol)$ pairs such that, 
for any $\e>0$, (\ref{eq:AR}) holds alongside
\bea
\label{eq:ASS1}
\Pr\{\Xol_J^n\ne \Xolhat_J^n)\} &\le& \e\\
\label{eq:ASS2}
\frac{1}{n} \E d_{ln}(\Xol_{J^c}^{n}, \Xolhat_{J^c}^{n}) &\le&  D_l + \e,
\quad l\in I_L.
\eea
The special case, where $M=2$, $J = \{1\}$ and $L=1$, was solved by Berger and Yeung \cite{BY}.

\end{enumerate}
Note that problem `T' is the same as problem `L' for $J=I_M$ and as problem `D' for $J = \{\}$.
Next we generalize the basic source coding problem to incorporate side information.

\subsection{Partial Side Information}
\label{sec:part}

First consider encoding of $\Xol$ using partial side information. Specifically, suppose $(\Xol,\Wol) \sim p(\xol,\wol)$ ($\Wol$ being a $K$-vector) and
draw $\{(\Xol(k),\Wol(k))\}$ i.i.d. $\sim p(\xol,\wol)$. Now encode $(\Xol,\Wol)$ using $M+K$ encoder mappings
\be
\label{eq:fmP}
\begin{array}{rcll} 
f_{m}&:&\Xcal_m^n \rightarrow \Zcal_m, \quad &m\in I_M\\
f_{M+m}&:&\Wcal_m^n \rightarrow \Zcal_{M+m}, \quad &m\in I_K
\end{array}
\ee
(i.e., $\fol:\Xcalol^{n}\times \Wcalol^n  \rightarrowtail \Zcalol$)
and decode using decoder mapping
$g: \Zcalol \rightarrow \Xcalol^n$ as in (\ref{eq:g}).
In other words, now estimate $\Xol^n$ by 
\be
\label{eq:rec'}
\Xolhat^n =g(\fol(\Xol^n,\Wol^n)).
\ee 
Note that only partial knowledge of side information $\Wol$ is available at the decoder ($\Wol^n$, however, is not estimated).
Further, a rate $(M+K)$-vector $\Rol$ is said to be achievable if for any $\e>0$, there exists (for $n$ sufficiently large) mapping pair $(\fol,g)$  such that 
\be
\label{eq:AR'1}
\frac{1}{n} \log |\Zcal_m | \le R_m +\e,\quad m\in I_{M+K}
\ee
and appropriate error or distortion criteria based on $(\Xol^n,\Xolhat^n)$ are met also within an accuracy of $\e$. 
Specifically, we modify the three basic source coding problems enumerated in Sec. \ref{sec:source} (each problem tag is now appended with `P') as follows.

We shall refer (\ref{eq:ALL})--(\ref{eq:ASS2}) below; in each case, assume
reconstruction $\Xolhat^n =g(\fol(\Xol^n,\Wol^n))$ as given in (\ref{eq:rec'}).
 
\begin{enumerate}

\item {\bf Lossless Coding (`LP'):}
The achievable set $\ALP$ is defined by the set of $\Rol$ such that, 
for any $\e>0$, (\ref{eq:AR'1}) holds alongside (\ref{eq:ALL}).
The special case, where $M=1$ and $K=1$, was solved by Wyner \cite{Wyner} and Ahlswede-K\"{o}rner \cite{AhlKor}.

\item {\bf Coding under Distortion Criteria (`DP'):}
The achievable set $\ADP$ is defined by the set of $(\Rol,\Dol)$ pairs such that, 
for any $\e>0$, (\ref{eq:AR'1}) holds alongside (\ref{eq:ALC}).
In the special case, where $M=1$, $K=1$ and $L=1$,
an inner bound on $\ADP$ was
found by Berger {\em et. al} \cite{Upper} and a complete solution was derived in our first paper \cite{PartI} of this series. 

\item {\bf Lossless Coding in Subset (`TP'):}
The achievable set $\ATP$ is defined by the set of $(\Rol,\Dol)$ pairs such that, 
for any $\e>0$, (\ref{eq:AR'1}) holds alongside
(\ref{eq:ASS1}) and
(\ref{eq:ASS2}).

\end{enumerate}
Of course, `P' is removed from any of the abovementioned tags if $\Wol$ is deterministic. 
Next we consider the case where additional side information is completely available at the decoder.

\subsection{Complete Side Information}
\label{sec:comp}

Suppose $(\Xol,\Wol,S) \sim p(\xol,\wol,s)$ ($S$ being a scalar) and
draw $\{(\Xol(k),\Wol(k),S(k))\}$ i.i.d. $\sim p(\xol,\wol,s)$. Now encode $(\Xol,\Wol)$ using $M+K$ encoder mappings
as in (\ref{eq:fmP}), i.e., $\fol:\Xcalol^{n}\times \Wcalol^n  \rightarrowtail \Zcalol$; however, decode using decoder mapping
\be
\label{eq:gC}
g: \Zcalol\times\Scal^n \rightarrow \Xcalol^n.
\ee
In other words, estimate $\Xol^n$ by 
\be
\label{eq:rec''}
\Xolhat^n =g( \fol(\Xol^n,\Wol^n),S^n).
\ee
Accordingly, we modify the four partial side information problems enumerated in Sec. \ref{sec:part} to also incorporate $S$ (each tag is further appended with `C') as follows. 

We refer (\ref{eq:ALL})--(\ref{eq:ASS2}) below; in each case, assume
reconstruction $\Xolhat^n =g( \fol(\Xol^n,\Wol^n),S^n)$ as given in (\ref{eq:rec''}).

\begin{enumerate}

\item {\bf Lossless Coding (`LPC'):}
The achievable set $\ALPC$ is defined by the set of $\Rol$ such that, 
for any $\e>0$, (\ref{eq:AR'1}) holds alongside (\ref{eq:ALL}).

\item {\bf Coding under Distortion Criteria (`DPC'):} 
The achievable set $\ADPC$ is defined by the set of $(\Rol,\Dol)$ pairs such that, 
for any $\e>0$, (\ref{eq:AR'1}) holds alongside (\ref{eq:ALC}).

\item {\bf Lossless Coding in Subset (`TPC'):} 
The achievable set $\ATPC$ is defined by the set of $(\Rol,\Dol)$ pairs such that, 
for any $\e>0$, (\ref{eq:AR'1}) holds alongside
(\ref{eq:ASS1}) and
(\ref{eq:ASS2}).

\end{enumerate}
Of course, `C' is removed from any of the abovementioned tags if $S$ is deterministic.  On the other hand, as seen in Sec. \ref{sec:part}, `P' is removed from any of the above tags, if $\Wol$ is deterministic. Correspondingly, problems `LC', `DC' and `TC' arise, where the only side information $S$ is completely available at the decoder. Note that Slepian-Wolf theorem solves
Problem `LC' completely \cite{SW}. Also,
the special case of problem `DC' (lossy coding with complete side information), where $M=1$ and $L=1$, was solved by Wyner and Ziv \cite{WZ}.

\subsection{Summary}
\label{sec:dep}

\begin{table}[t!]
\begin{center}
\begin{tabular}{|c|c|c|c|c|}
  \hline
  % after \\: \hline or \cline{col1-col2} \cline{col3-col4} ...
  &&{\bf Estimate}& {\bf Achievability} & {\bf Solution} \vspace*{-5pt}\\ 
  \rb{\bf Category} &\rb{\bf Tag} &{\bf (\boldmath{$\Xolhat^n$})} 
  &{\bf Conditions} &{\bf Status} \\
  \hline
%  \hline
  &`L'& $g( \fol(\Xol^n))$ & (\ref{eq:AR}), (\ref{eq:ALL})
  & Solved: General $M$.\\
  \raisebox{-0.5ex}[0pt]{Lossless} 
  &`LC'& $g(\fol(\Xol^n),S^n)$ & (\ref{eq:AR}), (\ref{eq:ALL})
  & Solved: General $M$.\\
  \raisebox{0.5ex}[0pt]{Coding} 
  &`LP'& $g( \fol(\Xol^n,\Wol^n))$ & (\ref{eq:AR'1}), (\ref{eq:ALL})
  & Solved: $M=1$, $K=1$.\\
  &`LPC'& $g( \fol(\Xol^n,\Wol^n),S^n)$ & (\ref{eq:AR'1}), (\ref{eq:ALL})
  & Unsolved.\\ 
  \hline
%  \hline
  &`D'& $g( \fol(\Xol^n))$ & (\ref{eq:AR}), (\ref{eq:ALC})
  & Solved: $M=1,2$, $L=1$.\\
  \raisebox{-1ex}[0pt]{Lossy} 
  &`DC'& $g(\fol(\Xol^n),S^n)$ & (\ref{eq:AR}), (\ref{eq:ALC})
  & Solved: $M=1$, $L=1$.\\
  \raisebox{0.1ex}[0pt]{Coding} 
    &&&& Solved:\\ 
  &\raisebox{2.7ex}[0pt]{`DP'}& \raisebox{2.7ex}[0pt]{$g( \fol(\Xol^n,\Wol^n))$} & \raisebox{2.7ex}[0pt]{(\ref{eq:AR'1}), (\ref{eq:ALC})}&\rb{$M=1$, $K=1$, $L=1$.} \vspace*{-8pt}\\
  &`DPC'& $g( \fol(\Xol^n,\Wol^n),S^n)$ & (\ref{eq:AR'1}), (\ref{eq:ALC})
  & Unsolved.\\ 
  \hline
%  \hline
  &&&& Solved:\\
  &\raisebox{2.7ex}[0pt]{`T'}& \raisebox{2.7ex}[0pt]{$g( \fol(\Xol^n))$} &
   \raisebox{2.7ex}[0pt]{(\ref{eq:AR}), (\ref{eq:ASS1}),
   (\ref{eq:ASS2})}
  & \rb{$M=2$, $J=\{1\}$, $L=1$.} \vspace*{-8pt}\\
  \raisebox{3ex}[0pt]{Lossless} 
  &`TC'& $g(\fol(\Xol^n),S^n)$ & (\ref{eq:AR}), (\ref{eq:ASS1}), (\ref{eq:ASS2})
  & Unsolved.\\
  \raisebox{4ex}[0pt]{Coding} 
  &`TP'& $g( \fol(\Xol^n,\Wol^n))$ & (\ref{eq:AR'1}), (\ref{eq:ASS1}), (\ref{eq:ASS2})
  & Unsolved.\\
  \raisebox{5ex}[0pt]{in Subset}
  &`TPC'& $g( \fol(\Xol^n,\Wol^n),S^n)$ & (\ref{eq:AR'1}), (\ref{eq:ASS1}), (\ref{eq:ASS2})
  & Unsolved.\\ 
  \hline
\end{tabular}
\end{center}
\caption{Summary and solution status of source coding problems.}
\label{tab:prob}
\end{table} 

So far we have identified twelve source coding problems in Secs. \ref{sec:source}, \ref{sec:part} and \ref{sec:comp}. In particular, we divided these problems into three categories: Lossless coding (`L$*$'), coding under distortion criterion (`D$*$') and lossless coding in subset (`T$*$'). 
Here `$*$' is one of blank, `P', `C' and `PC'. 
For quick reference, salient features of all the twelve problems are summarized in Table \ref{tab:prob}. In this paper, we solve all the abovementioned 
problems in their most general setting, 
save problems `L' and `LC', which are already completely solved.

Recall that problem `T$*$' reduces to problem `L$*$' if $J=I_M$ and to problem `D$*$' if $J = \{\}$ which we indicate by the diagram
\be
\label{eq:dep1}
\mbox{
\begin{tabular}{c}
{{~ `T$*$'}} \vspace{3pt}\\
{\small $J=I_M$}~{\Huge $\swarrow$} ~~ {\Huge $\searrow$} ~{\small $J=\{\}$}\vspace*{7pt} \\
{{`L$*$'}}~~~~~~~~~~~~{{`D$*$'}.}
\end{tabular}}
\ee
Also recall that the problem dependency due to side information
can be depicted by
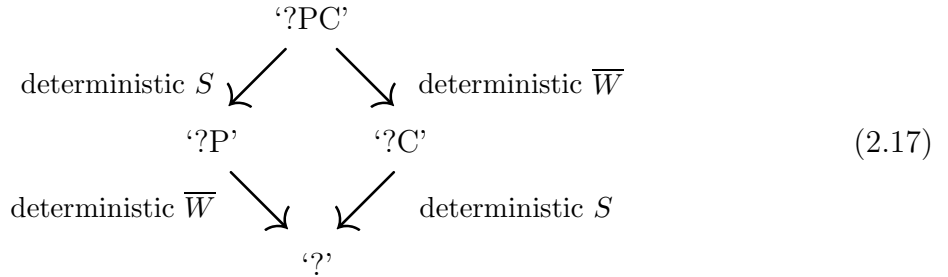
\be
\label{eq:dep2}
\mbox{
\begin{tabular}{c}
{{`?PC'}} \vspace{3pt}\\
~{\small deterministic $S$}~{\Huge $\swarrow$} ~~ {\Huge $\searrow$} ~{\small deterministic $\Wol$}\vspace*{7pt} \\
$\!\!${{`?P'}}~~~~~~~~~~~~~{{`?C'}} \vspace{3pt}\\
$\!\!$ $\!\!$ $\!\!${\small deterministic $\Wol$}~{\Huge $\searrow$} ~~ {\Huge $\swarrow$} ~{\small deterministic $S$}\vspace*{7pt} \\
{`?'}
\end{tabular}}
\ee
where `?' is one of `L', `D' and `T'. 
In view of the dependencies (\ref{eq:dep1}) and (\ref{eq:dep2}),
it is enough to solve problem `TPC' alone. The solution can then be specialized in order to solve other problems.

\Section{Unified Coding Theorem}
\label{sec:results}
We begin by giving a generic description of the solutions of all the twelve source coding problems. In particular, let $\Acal$ be the generic notation for the achievable rate or rate-distortion regions defined in Sec.~\ref{sec:prob}. Note that, since $\Acal$ is defined by appropriate $\e$--achievability conditions ($\e>0$), $\Acal$ is closed.
We state this in a unified coding theorem:

\begin{theorem}
\label{th:generic}
$\Acal = \AcalSol$. 
\end{theorem}

Here $\Acal^*= \bigcup_{n=1}^\infty \Acal_n^{*}$. We need to specify $\Acal_n^{*}$  
for each problem which we take up next.
In the process, we shall see that each $\Acal_n^{*}$ is closed. However, that does not necessarily imply $\Acal^*$ is closed. Hence the closure appears in Theorem \ref{th:generic}. 
In the following, we shall first specify $\AnTPC^*$ corresponding to problem `TPC', which we then specialize to the rest of the problems. 

\subsection{Lossless Coding in Subset}
\label{sec:TPC}

{\bf Problems `TPC' and `TP':} First consider problem `TPC'.
A rate-distortion pair $(\Rol\in\Rbb^{M+K},\Dol\in\Rbb^L)$  is said to belong to $\AnTPC^{*}$ if
there exist product of $(M+K)$ alphabets $\Zcalol$ (with the restriction $\Zcal_m=\Xcal_m^n$, $m\in J$),
conditional distributions $q_{m}(z_m|x_m^n)$, $m\in J^c$, and $r_{j}(z_{M+j}|w_j^n)$, $j\in I_K$, and mapping $\psi:\Zcalol \times \Scal^n \rightarrow\Xcalol^n_{J^c}$ such that
\bea
\label{eq:R1I.}
\onen
I(\Xol_I^n;\Zol_I|\Zol_{I_{M+K}\setminus I},S^n) &\le& R_I, \quad I\subseteq I_M\setminus \{\}\\
\label{eq:R2I.}
\onen I(\Wol_I^n;\Zol_{M+I}|\Zol_{M+ I_K\setminus I},S^n) &\le& R_{M+I}, \quad I\subseteq I_K\setminus \{\}\\
\label{eq:D.}
\onen d_{ln}(\Xol^n_{J^c},\psi(\Zol,S^n))&\le& D_l,\quad l\in I_L
\eea
where $\Zol_J=\Xol_J^n$,
\be
\label{eq:pTPC}
(\Xol^n,\Wol^n,S^n,\Zol_{I_{M+K}\setminus J}) \sim p_n(\xol^n,\wol^n,s^n)
\prod_{m\in J^c} q_{m}(z_m|x_m^n)
\prod_{j\in I_K} r_{j}(z_{M+j}|w_j^n)
\ee 
and $p_n(\xol^n,\wol^n,s^n) = \prod_{k=1}^n p(\xol(k),\wol(k),s(k))$. Denote $\Xol' = (\Xol,\Wol)$ such that $\Xol'_{I_M} = \Xol$ and $\Xol'_{M+I_K} = \Wol$. Then, by (\ref{eq:pTPC}), $U \rightarrow {\Xol'_I}^n\rightarrow\Zol_I$
forms Markov chain for any $I\subseteq I_{M+K}$ and any subcollection $U$ of $({\Xol'}^n,\Zol,S^n)$ excluding $({\Xol'_I}^n,\Zol_I)$.
Further, we show in Appendix \ref{app:TPC} that, splitting each $I=I'\cup I''\subseteq I_M\setminus \{\}$ such that $I'\subseteq J$ and $I''\subseteq J^c$, we can equivalently write (\ref{eq:R1I.}) as
\be
\label{eq:TPC.}
\onen H(\Xol_{I'}^n|\Xol^n_{J\setminus I'}, 
\Zol_{I_{M+K}\setminus(J\cup I'')},S^n)
+
\onen I(\Xol_{I''}^n;\Zol_{I''}|\Xol^n_{J}, 
\Zol_{I_{M+K}\setminus(J\cup I'')},S^n)\le R_{I'}+R_{I''}.
\ee
Note that the total number of conditions given by (\ref{eq:R1I.}) (or, equivalently, (\ref{eq:TPC.}))   and (\ref{eq:R2I.}) is $(2^{M}+2^{K}-2)$. Further, consider Problem `TP' and define $\AnTP^*=\AnTPC^*$ such that $S$ is deterministic in (\ref{eq:R1I.})--(\ref{eq:pTPC}) (or, in (\ref{eq:TPC.}) instead of (\ref{eq:R1I.})), i.e., occurrences of $S^n$ (and $s^n$)
are simply removed.

\noindent {\bf Problems `TC' and `T':} 
First consider problem `TC' and define $\AnTC^*=\AnTPC^*$ such that $\Wol$ is deterministic in (\ref{eq:R1I.})--(\ref{eq:pTPC}).
Note that the left hand side in (\ref{eq:R2I.}) is now zero, i.e., it is enough to consider only $\Rol = (R_1,R_2,...,R_M)\in \Rbb^M$. 
Also, by (\ref{eq:pTPC}), $\Zol_{M+I_K}$ is independent of $(\Xol^n,S^n,\Zol_{I_M})$, hence occurrences of components of $\Zol_{M+I_K}$ can be removed from (\ref{eq:R1I.}), (\ref{eq:D.}) and (\ref{eq:pTPC}). 
Writing afresh, 
a rate-distortion pair $(\Rol\in\Rbb^{M},\Dol\in\Rbb^L)$ belongs to $\AnTC^{*}$ if
there exist product of $M$ alphabets $\Zcalol$ (now playing the role of the abovementioned $\Zcalol_{I_M}$),
conditional distributions $q_{m}(z_m|x_m^n)$, $m\in I_M$,  and mapping $\psi:\Zcalol \times \Scal^n \rightarrow\Xcalol^n_{J^c}$ such that
\bea
\label{eq:R1I.3}
\onen
I(\Xol_I^n;\Zol_I|\Zol_{I^c},S^n) &\le& R_I, \quad I\subseteq I_M\setminus \{\}\\
\label{eq:D.3}
\onen d_{ln}(\Xol^n_{J^c},\psi(\Zol,S^n))&\le& D_l,\quad l\in I_L
\eea
where $\Zol_J=\Xol_J^n$ and
$
(\Xol^n,S^n,\Zol_{J^c}) \sim p_n(\xol^n,s^n)
\prod_{m\in J^c} q_{m}(z_m|x_m^n)$. 
Referring to (\ref{eq:TPC.}), (\ref{eq:R1I.3}) can equivalently be written as
\be
\label{eq:TPC..}
\onen H(\Xol_{I'}^n|\Xol^n_{J\setminus I'}, 
\Zol_{J^c\setminus I''},S^n)
+
\onen I(\Xol_{I''}^n;\Zol_{I''}|\Xol^n_{J}, 
\Zol_{J^c\setminus I''},S^n)\le R_{I'}+R_{I''}
\ee
where, as earlier, we split $I=I'\cup I''\subseteq I_M\setminus \{\}$ such that $I'\subseteq J$ and $I''\subseteq J^c$.
Further, consider Problem `T' and define $\AnT^*=\AnTC^*$ such that $S$ is deterministic in (\ref{eq:R1I.3}) and (\ref{eq:D.3}) (or, in (\ref{eq:TPC..}) instead of (\ref{eq:R1I.3})), i.e., occurrences of $S^n$ (and $s^n$)
are simply removed. Note that, in the special case, where $M=2$, $J=\{1\}$, $L=1$ and $S^n$ is deterministic, (\ref{eq:TPC..}) and (\ref{eq:D.3}) are the same as conditions (6.28)--(6.31) of \cite{PartI}, which define $\Acal_n^*$ for Berger-Yeung problem \cite{BY}.

\subsection{Lossless Coding}

{\bf Problems `LPC' and `LP':} Now consider problem `LPC', which is problem `TPC' with $J=I_M$. In this case, (\ref{eq:TPC.}) takes the form
$$\onen H(\Xol_{I}^n|\Xol^n_{I^c}, 
\Zol_{M+I_{K}},S^n)\le R_I$$ because now $I'=I\subseteq I_M\setminus \{\}$, $I''=\{\}$ and $I_{M+K}\setminus (J\cup I'') = M+ I_K$. Of course, (\ref{eq:D.}) does not arise because
distortion criteria $d_l$'s are no longer defined.
Hence, writing (\ref{eq:TPC.}), (\ref{eq:R2I.}) and (\ref{eq:pTPC}) afresh, 
a rate vector $\Rol\in\Rbb^{M+K}$ belongs to $\AnLPC^{*}$ if
there exist product of $K$ alphabets $\Zcalol$ (now playing the role of abovementioned $\Zol_{M+I_K}$) and
conditional distributions $r_{j}(z_j|w_j^n)$, $j\in I_K$, such that
\bea
\label{eq:R1I}
\onen
H(\Xol_I^n|\Xol^n_{I^c},\Zol,S^n) &\le& R_I, \quad I\subseteq I_M\setminus \{\}\\
\label{eq:R2I}
\onen I(\Wol_I^n;\Zol_{I}|\Zol_{ I^c},S^n) &\le& R_{M+I}, \quad I\subseteq I_K \setminus \{\}
\eea
where 
\be
\label{eq:pLLPC}
(\Xol^n,\Wol^n,S^n,\Zol) \sim p_n(\xol^n,\wol^n,s^n)
\prod_{j\in I_K} r_{j}(z_j|w_j^n)
\ee 
and $p_n(\xol^n,\wol^n,s^n) = \prod_{k=1}^n p(\xol(k),\wol(k),s(k))$. Note that, for any $I\subseteq I_M$ and $I'\subseteq I_K$,  $(\Xol_{I}^n,S^n)\rightarrow\Wol_{I'}^n\rightarrow\Zol_{I'}$ is Markov chain. Also note that the total number of conditions given by (\ref{eq:R1I}) and (\ref{eq:R2I}) is $(2^{M}+2^{K}-2)$. 
Further, consider Problem `LP' and define $\AnLP^*=\AnLPC^*$ such that $S$ is deterministic in (\ref{eq:R1I})--(\ref{eq:pLLPC}), i.e., occurrences of $S^n$ (and $s^n$)
are simply removed. Note that, in the special case, where $M=1$, $K=1$ and $S^n$ is deterministic, (\ref{eq:R1I}) and (\ref{eq:R2I}) are the same as conditions (6.14) and (6.15) of \cite{PartI}, which define $\Acal_n^*$ for the so-called ``side information problem'' \cite{Wyner,AhlKor}.

\noindent {\bf Problems `LC' and `L':} Consider problem `LC' and define $\AnLC^*=\AnLPC^*$ such that $\Wol$ is deterministic in (\ref{eq:R1I})--(\ref{eq:pLLPC}).
Note that the left hand side in (\ref{eq:R2I}) is zero, i.e., it is enough to consider only $\Rol = (R_1,R_2,...,R_M)\in \Rbb^M$. Also, by (\ref{eq:pLLPC}), $\Zol$ in independent of $(\Xol^n,S^n)$, hence $\Zol$ can be removed from (\ref{eq:R1I}), i.e., we have 
\be
\label{eq:R1I'}
\onen
H(\Xol_I^n|\Xol^n_{I^c},S^n) 
= H(\Xol_I|\Xol_{I^c},S)
\le R_I, \quad I\subseteq I_M\setminus \{\}.
\ee
Hence, observe that $\AnLC^*=\AoneLC^*$, i.e., $\ALCol= \ALC^* = \AoneLC^*$. In view of this, Theorem \ref{th:generic} is a version of Slepian-Wolf theorem \cite{Cover}. Further, consider Problem `L' and define $\AnL^* = \AoneL^* = \AoneLC^*$ with deterministic $S$ in (\ref{eq:R1I'}). In this case, Theorem \ref{th:generic} is the usual statement of Slepian-Wolf theorem.

\subsection{Coding under Distortion Criteria}

{\bf Problems `DPC' and `DP':} Next consider problem `DPC', which is problem `TPC' with $J=\{\}$. Rewriting (\ref{eq:R1I.})--(\ref{eq:pTPC}) for this special case,
a rate-distortion pair $(\Rol\in\Rbb^{M+K},\Dol\in\Rbb^L)$ belongs to $\AnDPC^{*}$ if
there exist product of $(M+K)$ alphabets $\Zcalol$,
conditional distributions $q_{m}(z_m|x_m^n)$, $m\in I_M$, and $r_{j}(z_{M+j}|w_j^n)$, $j\in I_K$, and mapping $\psi:\Zcalol \times \Scal^n \rightarrow\Xcalol^n$ such that
\bea
\label{eq:'R1I}
\onen
I(\Xol_I^n;\Zol_I|\Zol_{I_{M+K}\setminus I},S^n) &\le& R_I, \quad I\subseteq I_M\setminus \{\}\\
\label{eq:'R2I}
\onen I(\Wol_I^n;\Zol_{M+I}|\Zol_{M+ I_K\setminus I},S^n) &\le& R_{M+I}, \quad I\subseteq I_K\setminus \{\}\\
\label{eq:'D}
\onen d_{ln}(\Xol^n,\psi(\Zol,S^n))&\le& D_l,\quad l\in I_L
\eea
where 
\be
\label{eq:pLCPC}
(\Xol^n,\Wol^n,S^n,\Zol) \sim p_n(\xol^n,\wol^n,s^n)
\prod_{m\in I_M} q_{m}(z_m|x_m^n)
\prod_{j\in I_K} r_{j}(z_{M+j}|w_j^n)
\ee 
and $p_n(\xol^n,\wol^n,s^n) = \prod_{k=1}^n p(\xol(k),\wol(k),s(k))$.  
Again note that the total number of conditions given by (\ref{eq:'R1I}) and (\ref{eq:'R2I}) is $(2^{M}+2^{K}-2)$. Further, consider Problem `DP' and define $\AnDP^*=\AnDPC^*$ such that $S$ is deterministic in (\ref{eq:'R1I})--(\ref{eq:pLCPC}), i.e., occurrences of $S^n$ (and $s^n$)
are simply removed. 

\noindent {\bf Problems `DC' and `D':} 
Consider problem `DC' and define $\AnDC^*=\AnDPC^*$ such that $\Wol$ is deterministic in (\ref{eq:'R1I})--(\ref{eq:pLCPC}).
Note that the left hand side in (\ref{eq:'R2I}) is zero, i.e., it is enough to consider only $\Rol = (R_1,R_2,...,R_M)\in \Rbb^M$. 
Also, by (\ref{eq:pLCPC}), $\Zol_{M+I_K}$ in independent of $(\Xol^n,S^n,\Zol_{I_M})$, hence components of $\Zol_{M+I_K}$ can be removed from (\ref{eq:'R1I}), (\ref{eq:'D}) and (\ref{eq:pLCPC}). Writing afresh, 
A rate-distortion pair $(\Rol\in\Rbb^{M},\Dol\in\Rbb^L)$ belongs to $\AnDC^{*}$ if
there exist product of $M$ alphabets $\Zcalol$ (now playing the role of the abovementioned $\Zcalol_{I_M}$),
conditional distributions $q_{m}(z_m|x_m^n)$, $m\in I_M$,  and mapping $\psi:\Zcalol\times \Scal^n \rightarrow\Xcalol^n$ such that
\bea
\label{eq:''R1I}
\onen
I(\Xol_I^n;\Zol_I|\Zol_{I^c},S^n) &\le& R_I, \quad I\subseteq I_M\\
\label{eq:''D}
\onen d_{ln}(\Xol^n,\psi(\Zol,S^n))&\le& D_l,\quad l\in I_L
\eea
where 
\be
\label{eq:pDC}
(\Xol^n,\Wol^n,S^n,\Zol) \sim p_n(\xol^n,\wol^n,s^n)
\prod_{m\in I_M} q_{m}(z_m|x_m^n).
\ee 
Further, consider Problem `D' and define $\AnD^*=\AnDC^*$ such that $S$ is deterministic in (\ref{eq:''R1I})--(\ref{eq:pDC}), i.e., occurrences of $S^n$ (and $s^n$)
are simply removed. 

It is enough to prove Theorem \ref{th:generic} for problem `TPC', which, as we have just seen, specializes to Theorem \ref{th:generic} for each of the rest of the problems at hand. We present the proof 
in Sec.~\ref{sec:proofM}, which requires a fundamental principle of multiterminal source coding that generalizes our earlier results given in Theorem 3.1 and Lemma 3.5 of \cite{PartI}. We first state and prove this generalized principle (Theorem \ref{th:funda}) in Sec.~\ref{sec:funda}.

\Section{Fundamental Principle}
\label{sec:funda}
\subsection{Statement}

\begin{theorem}
\label{th:funda}
Let $(\Yol,\Zol,V)= (Y_1,Y_2,...,Y_{M'}, Z_1,Z_2,...,Z_{M'},V)$ be a collection of ${2{M'}+1}$ random variables $\sim p'(\yol,v)\prod_{m=1}^{M'} q'_{m}(z_m|y_m)$ and let $\{(\Yol(k),V(k))\}$
be i.i.d. copies of $(\Yol,V)$. 
Then for any rate $M'$-vector $\Rol'$ such that
\be
\label{eq:Rdef}
I\left(\Yol_{I};\Zol_I|\Zol_{I^c},V\right) \le R'_I
\ee
for all $I\subseteq I_{M'}\setminus \{\}$ and for any $\e' \rightarrow 0$, there exists a sequence of mapping pairs $(\fol:\Ycalol^{n'} \rightarrowtail \Ucalol,g:\Ucalol\times\Vcal \rightarrow \Zcalol^{n'})$ for some sequence $\Ucalol$ of $M'$-fold product of alphabets 
(and some $n'\rightarrow \infty$)
 such that 
\bea
\label{eq:rate}
\frac{1}{n'} \log |\Ucal_i| &\le& R'_i + \e'',\quad i\in I_{M'}\\
\label{eq:prob}
\Pr\{\Ecal\} &\le& \e''
\eea
where 
$$\Ecal =\{(\Yol^{n'},
\Zolhat^{n'},V^{n'})\notin \Tenn(\Yol,\Zol,V)\},$$
$\Zolhat^{n'}=g(\fol(\Yol^{n'}),V^{n'})$ and $\e''\rightarrow 0$.
\end{theorem}

Here note that $f_i:\Ycal_i^{n'}\rightarrow\Ucal_i$, $i\in I_{M'}$, i.e., encoders do not cooperate. 
Also
note that $U \rightarrow {\Yol_I}\rightarrow\Zol_I$
forms Markov chain for any $I\subseteq I_{M'}$ and any subcollection $U$ of $(\Yol,\Zol,V)$ excluding $({\Yol_I},\Zol_I)$.
Further, due to strong law of large numbers, in the above $(\Yol^{n'},V^{n'})$ can be replaced, without loss of generality, by any $(\widehat{\Yol}^{n'},\widehat{V}^{n'})$ such that $\Pr\{(\widehat{\Yol}^{n'},\widehat{V}^{n'})
\notin \Tenn(\Yol,V)\}\le \e''_1$, where $\e''_1\rightarrow 0$ as $\e'\rightarrow 0$. Such substitutions are standard and will sometimes be carried out without explicit mention. Also,
note that as $\e'\rightarrow 0$, $n'\rightarrow\infty$ through values $n'>n'_0(\e')$ for appropriate $n'_0(\cdot)$. 

Theorem \ref{th:funda} roughly states the following. Using a sequence of codes $(\fol,g)$ (of sufficiently large length $n'$), one can achieve any rate $M'$-vector $\Rol'$ satisfying the $2^{M'}-1$ inequalities given by (\ref{eq:Rdef}) such that the estimate $\Zolhat^{n'}=g(\fol(\Yol^{n'}),V^{n'})$ of $\Zol^{n'}$, based on the encoding $\fol(\Yol^{n'})$ and side information $V^{n'}$,
is jointly typical with $\Yol^{n'}$ with high probability. 
Further, Theorem \ref{th:funda} includes Slepian--Wolf's direct theorem as a special case. To see this, set
$\Zol=\Yol$ and let $V$ be deterministic, so that $I\left(\Yol_{I};\Zol_I|\Zol_{I^c},V\right) = H(\Yol_I|\Yol_{I^c})$ in (\ref{eq:Rdef}). Hence, for any rate vector $\Rol'$ such $H\left(\Yol_{I}|\Yol_{I^c}\right) \le R'_I$, $I\subseteq I_{M'}\setminus\{\}$, (\ref{eq:rate}) holds and
$\Pr\{\Yol^{n'}\ne \Yolhat^{n'}\}$ is arbitrarily small (due to (\ref{eq:prob})). Next we turn to the proof of Theorem \ref{th:funda}.

\subsection{Necessary Ingredients}
\label{sec:ingre}

Before proceeding any further, let us point out that we derived in Lemma 3.5 of our earlier work \cite{PartI}
a special case of Theorem \ref{th:funda} where $M'=2$ and $(Z_2,V)$ is deterministic. In fact,
this special case was demonstrated to encapsulate the essence of Wyner-Ziv's typicality argument.
In other words, proving Theorem \ref{th:funda} amounts to generalizing 
an earlier result which we reproduce below for ease of reference.

\begin{lemma}
\label{le:adi2}
\cite[Lemma 3.5]{PartI}
Let $(Y_1,Y_2,Z_1)\sim p'(y_1,y_2)q'_1(z_1|y_1)$ and draw $\{(Y_{1}(k),Y_{2}(k))\}$ i.i.d. $\sim p'(y_1,y_2)$.
Then for 
any rate $R'_1$ such that
\be
\label{eq:Rdef''}
I(Y_1;Z_1|Y_2)\le R'_1
\ee
and any $\e'\rightarrow 0$,
there exists a sequence of 
mapping pairs
$$(f_1:\Ycal_1^{n'} \rightarrow \Ucal_1,g:\Ucal_1\times \Ycal_2^{n'} \rightarrow \Zcal_1^{n'})$$
for some sequence of alphabets $\Ucal_1$ (and some $n'\rightarrow\infty$) 
such that
\bea
\label{eq:ShRe'}
\frac{1}{n'} \log |\Ucal_1| &\le& R'_1 + \epsilon''\\
\label{eq:ShPr'}
\Pr\{(Y_1^{n'},Y_2^{n'},\Zhat_1^{n'}) \notin \Tenn(Y_1,Y_2,Z_1)\} &\le& \epsilon''
\eea
where 
$\Zhat_1^{n'} = g(f_1(Y_1^{n'}),Y_2^{n'})$
 and $\e''\rightarrow 0$.
\end{lemma}

Note that $Y_2 \rightarrow Y_1 \rightarrow Z_1$ is required to form Markov chain for Lemma \ref{le:adi2} to apply. Further, we shall require another crucial result. In particular, denote by $\Bcal^*$
the convex rate region  defined by the conditions given in (\ref{eq:Rdef}).
We need to exhaustively identify the corner points of $\Bcal^*$, which are rates $\Rol'$ such that $M'$ of the $2^{M'}-1$ constraints in (\ref{eq:Rdef}) are active.  
For this purpose, we shall make the mild assumption that the random variables $(\Yol,\Zol,V)$ are all dependent. Under this assumption, any Markov property of any subset of $(\Yol,\Zol,V)$ does not hold if such property does not follow directly from the generic definition $(\Yol,\Zol,V)\sim p'(\yol,v)\prod_{m=1}^{M'} q'_{m}(u_m|y_m)$.

\begin{lemma}
\label{le:corner}
Let $\Pi$ be the set of permutations of $(1,2,...,M')$. Further, denote by $\pi(i)$ the permuted position of $i\in I_{M'}$ under permutation $\pi\in \Pi$. Also denote $\pi(I) = \{\pi(i):i\in I\}$. 
Then
$\Bcal^*$ has $M'!$ corner points $\Rol^{\prime * (\pi)}$ indexed by $\pi\in\Pi$ such that
\be
\label{eq:Rpi}
{R^{\prime * (\pi)}_{\pi(i)}} = I(Y_{\pi(i)};Z_{\pi(i)}|\Zol_{\pi(I_{i-1})},V), \quad i\in I_{M'}.
\ee
\end{lemma}

The proof of Lemma \ref{le:corner} is somewhat involved and is relegated 
to
 Appendix \ref{sec:cornerMom}.
Corresponding to the identity permutation $\pi_1$, from (\ref{eq:Rpi}), we have
\be
\label{eq:cor1}
\Rol^{\prime *(\pi_1)} = (I(Y_1;Z_1|V), ~I(Y_2;Z_2|Z_1,V),~...,~I(Y_{M'};Z_{M'}|\Zol_{I_{M'-1}},V))
\ee
as we shall also see in Lemma \ref{le:cornQ}. Further, by Lemma \ref{le:corner} and referring to (\ref{eq:Rdef}), any $\Rol'\in \Bcal^*$ can be written as
\be
\label{eq:convex}
\sum_{\pi\in \Pi} \lambda_\pi R_i^{\prime * (\pi)} \le R'_i, \quad i\in I_{M'}
\ee
for some $\{\lambda_\pi\}_{\pi\in\Pi}$ such that each $\lambda_\pi \ge 0$ and $\sum_{\pi\in \Pi} \lambda_\pi =1$.
For our analysis,
we shall also require the Markov lemma. In the following 
we give a version that
rewords Lemma 14.8.1 of \cite{Cover} and appears in its present form in Lemma 3.7 of \cite{PartI}. 

\begin{lemma}
\label{le:Mar} 
\cite[Lemma 14.8.1]{Cover} \cite[Lemma 3.7]{PartI}
Let $(Y_1,Y_2,Z_1)\sim p'(y_1,y_2)q'_1(z_1|y_1)$ and the sequence 
of triplets 
$\{(\hat{Y}_1(k),\hat{Y}_2(k),\hat{Z}_1(k))\}$
be such that, for any $\e'\rightarrow 0$ (and appropriate $n'\rightarrow \infty$), 
\beann
\Pr\{(\hat{Y}_1^{n'},\hat{Y}_2^{n'})\notin\Tenn(Y_1,Y_2)\}&\le& \e'_1\\ 
\Pr\{\hat{Y}_1^{n'},\hat{Z}_1^{n'})\notin\Tenn(Y_1,Z_1)\} &\le& \e'_1
\eeann 
for some $\e'_1\rightarrow 0$. Then $$\Pr\{(\hat{Y}_1^{n'},\hat{Y}_2^{n'},\hat{Z}_1^{n'})
\notin\Tenn(Y_1,Y_2,Z_1)\}\le \e'_2$$ 
for some $\e'_2$ such that $\e'_2\rightarrow 0$ as $\e'\rightarrow 0$. 
\end{lemma}

Now we are ready to complete the proof of Theorem \ref{th:funda}.

\subsection{Proof of Theorem \ref{th:funda}}
\label{sec:proof}

Denote by $\Bcal$ the set
of rate vectors $\Rol'$ such that for any $\e' \rightarrow 0$ there exists a sequence of mapping pairs $(\fol,g)$ satisfying (\ref{eq:rate}) and (\ref{eq:prob}). 
Clearly, Theorem \ref{th:funda} states 
$\Bcal^*\subseteq \Bcal$. 
Next
we claim $\Bcal$ is convex.
To see this, note that
if each of two rate vectors ${\Rol}^{\prime(0)}$ and ${\Rol}^{\prime(1)}$ belongs to $\Bcal$, then, by appropriate time sharing, we can ensure that any convex combination of such vectors also belongs to $\Bcal$. 

\begin{lemma}
\label{le:ulti} Any $\Rol'$, such that 
\be
\label{eq:cor3}
R^{\prime *(\pi_1)}_{i} = I(Y_{i};Z_{i}|\Zol_{I_{i-1}},V) \le R'_i,\quad i\in I_{M'}
\ee
($\pi_1$ being identity permutation),
belongs to $\Bcal$.
\end{lemma}

{\bf {\em Proof}:} Let us begin by 
noting that condition (\ref{eq:cor3}) for $i=1$ is same as condition (\ref{eq:Rdef''}) with $(Y_1,V,Z_1)$ playing the role of $(Y_1,Y_2,Z_1)$ ($V\rightarrow Y_1 \rightarrow Z_1$, of course, forms Markov chain). Hence, by Lemma \ref{le:adi2}, for any $\e'\rightarrow 0$, there exists a sequence of mapping pairs $(f_1: \Ycal_1^{n'} \rightarrow \Ucal_1, g_1:\Ucal_1\times\Vcal^{n'}\rightarrow \Zcal_1^{n'})$ such that
\bea
\label{eq:SR}
\frac{1}{n'} \log |\Ucal_1| &\le& R'_1 + \epsilon'_1\\
\label{eq:SP}
\Pr\{\Ecal_1\} &\le& \epsilon'_1
\eea
where $$\Ecal_1=\{(Y_1^{n'},\Zhat_1^{n'},V^{n'}) \notin \Tenn(Y_1,Z_1,V)\},$$
$\Zhat_1^{n'} = g_1(f_1(Y_1^{n'}),V^{n'})$
 and $\e'_1\rightarrow 0$.
 
 In fact, for any $\e' \rightarrow 0$, 
 next we show that 
 there exists a sequence of mapping pairs $(f_i: \Ycal_i^{n'} \rightarrow \Ucal_i, g_i:\Ucal_i\times(\Zcalol_{I_{i-1}}^{n'}\times\Vcal^{n'})\rightarrow \Zcal_i^{n'})$ for each $i\in I_{M'}$ such that
\bea
\label{eq:SRi}
\frac{1}{n'} \log |\Ucal_i| &\le& R'_i + \epsilon'_i\\
\label{eq:SPi}
\Pr\{\Ecal_i\} &\le& \epsilon'_i
\eea
where $$\Ecal_i=\{(\Yol_{I_i}^{n'},\Zolhat^{n'}_{I_{i}},V^{n'}) \notin \Tenn(\Yol_{I_i},\Zol_{I_{i}},V)\},$$
$\Zhat_i^{n'} = g_i(f_i(Y_i^{n'}),(\Zolhat^{n'}_{I_{i-1}},V^{n'}))$
 and $\e'_i\rightarrow 0$. We have already seen that the above result holds for $i=1$. We shall show this for general $i\in I_{M'}$ by induction. Specifically, we assume that the result holds for $i-1$ (in place of $i$) for some $i\in\{2,3,...,M'\}$. It is enough to show the result for $i$ under the above assumption. 
  
First, writing $i-1$ in place of $i$ in (\ref{eq:SPi}), we have
\be
 \label{eq:SP..i}
\Pr\{(\Yol_{I_{i-1}}^{n'},\Zolhat_{I_{i-1}}^{n'},V^{n'}) \notin \Tenn(\Yol_{I_{i-1}},\Zol_{I_{i-1}},V)\} \le \epsilon'_{i-1}
\ee 
where $\epsilon'_{i-1} \rightarrow 0$ as $\e' \rightarrow 0$.
Further, noting that $Y_i\rightarrow(\Yol_{I_{i-1}},V) \rightarrow \Zol_{I_{i-1}}$ forms Markov chain and
using Lemma \ref{le:Mar},  we have, from (\ref{eq:SP..i}),
\be
 \label{eq:SP'..i}
\Pr\{(\Yol_{I_{i}}^{n'},\Zolhat_{I_{i-1}}^{n'},V^{n'}) \notin \Tenn(\Yol_{I_{i}},\Zol_{I_{i-1}},V)\} \le \epsilon''_{i-1}
\ee  
 where $\epsilon''_{i-1} \rightarrow 0$ as $\e' \rightarrow 0$. Hence, of course,
\be
 \label{eq:SP'..i.}
\Pr\{(Y_i^{n'},\Zolhat_{I_{i-1}}^{n'},V^{n'}) \notin \Tenn(Y_i,\Zol_{I_{i-1}},V)\} \le \epsilon''_{i-1}.
\ee  
Also
note that condition (\ref{eq:cor3}) is same as condition (\ref{eq:Rdef''}) with $(Y_i,(\Zol_{I_{i-1}},V),Z_i)$ in place of $(Y_1,Y_2,Z_1)$ ($(\Zol_{I_{i-1}},V)\rightarrow Y_i \rightarrow Z_i$, of course, forms Markov chain). 
As a result, by Lemma \ref{le:adi2}, there exists a sequence of mapping pairs $(f_i, g_i)$ such that 
\bea
\label{eq:SRi'}
\frac{1}{n'} \log |\Ucal_i| &\le& R'_i + \epsilon'_i\\
\label{eq:SPi'}
\Pr\{(Y_i^{n'},\Zhat_i^{n'},\Zolhat_{I_{i-1}}^{n'},V^{n'}) \notin \Tenn(Y_i,Z_i,\Zol_{I_{i-1}},V)\} &\le& \epsilon'_i
\eea
where
$\Zhat_i^{n'} = g_i(f_i(Y_2^{n'}),(\Zolhat_{I_{i-1}}^{n'},V^{n'}))$
 and $\e'_i\rightarrow 0$. Further, noting the subtle fact that $\Yol_{I_{i-1}}\rightarrow(Y_i,\Zol_{I_{i-1}},V) \rightarrow Z_i$ forms Markov chain and
using Lemma \ref{le:Mar}, we have, from (\ref{eq:SP'..i}) and (\ref{eq:SPi'}),
\bea
\label{eq:SPi''}
\Pr\{(\Yol_{I_i}^{n'},\Zolhat_{I_i}^{n'},V^{n'}) \notin \Tenn(\Yol_{I_i},\Zol_{I_i},V)\} &\le& \epsilon''_i
\eea 
where $\e''_i\rightarrow 0$. Hence (\ref{eq:SRi'}) and (\ref{eq:SPi''}) give  (\ref{eq:SRi}) and (\ref{eq:SPi}), respectively, with  $\max\{\e'_i,\e''_i\}$ now playing the role of $\e'_i$.

Recall that we have $\fol:\Ycalol^{n'} \rightarrowtail \Ucalol$. Further, 
recalling
\beann
\Zhat_1^{n'} &=& g_1(f_1(Y_1^{n'}),V^{n'})\\
\Zhat_2^{n'} &=& g_2(f_2(Y_2^{n'}),\Zhat_1^{n'},V^{n'})\\
&=& g_2(f_2(Y_2^{n'}),g_1(f_1(Y_1^{n'}),V^{n'}),V^{n'})\\
&\vdots&
\eeann
and so on, we have
$$
\Zolhat^{n'} = g(\fol(\Yol^{n'}),V^{n'})
$$
for certain $g:\Ucalol\times\Vcal^{n'} \rightarrow \Zcalol^{n'}$. Thus, for the mapping pair $(\fol,g)$, (\ref{eq:SRi}) and (\ref{eq:SPi}) hold for each $i\in I_{M'}$. Writing $\e'' = \max_{i} \{\e'_i\}$ and collecting condition (\ref{eq:SRi}) for each $i\in I_{M'}$, we get (\ref{eq:rate}). Further, for $i=M'$, (\ref{eq:SPi}) is (\ref{eq:prob}). Hence $\Rol'\in \Bcal$. \hfill$\Box$

Now we are ready to complete the proof of Theorem \ref{th:funda} by showing $\Bcal^*\subseteq \Bcal$.

{\bf {\em Proof of Theorem \ref{th:funda}}:} By Lemma \ref{le:ulti},
any $\Rol^{\prime}$, such that 
$R^{\prime * (\pi_1)}_{i} \le R^{\prime}_i$, $i\in I_{M'}$, belongs to $\Bcal$. Generalizing this by symmetry, 
any $\Rol^{\prime(\pi)}$, such that 
$R^{\prime *(\pi)}_{i} \le R^{\prime(\pi)}_i$, $i\in I_{M'}$, for arbitrary permutation $\pi\in \Pi$, belongs to $\Bcal$. Further, by (\ref{eq:convex}), any $\Rol'\in \Bcal^*$ can be written as a convex combination of at most $M'!$ such $\Rol^{\prime(\pi)}$'s. Since each such $\Rol^{\prime(\pi)}$ also belongs to $\Bcal$, we have $\Rol'\in \Bcal$ due to convexity of $\Bcal$. Hence $\Bcal^*\subseteq \Bcal$. \hfill$\Box$  

Recall that we assumed statistical dependence of the random variables $(\Yol,\Zol,V)$ in the above proof.
However, Theorem \ref{th:funda} holds even when admissible subsets of $(\Yol,\Zol,V)$ are independent. In such case, some of the constraints given in (\ref{eq:Rdef}) degenerate. However, one can still identify the desired corner points of the resulting $\Bcal^*$ (which of course remains convex) and prove Theorem \ref{th:funda} mimicking our analysis. 

\Section{Proof of Theorem \ref{th:generic} for Problem `TPC'}
\label{sec:proofM}
At this point let us turn to Theorem \ref{th:generic} ($\Acal=\AcalSol$), which we shall prove for problem `TPC'. For the sake of convenience, we shall drop the subscript `TPC' throughout this section. The proof consists of two parts:
The inner bound $\Acal \supseteq \AcalSol$ is shown in Sec. \ref{sec:2In} using the fundamental principle given in Theorem \ref{th:funda}. The outer bound $\Acal \subseteq \AcalSol$ is shown in Sec. \ref{sec:2Out} with the aid of Slepian-Wolf theorem \cite{SW} and Fano's inequality \cite{Cover}. For ease of reference, we first reproduce in Sec.~\ref{sec:rep} the definitions of $\Acal$ and $\Acal^*$ for problem `TPC' from Secs.~\ref{sec:comp} and \ref{sec:TPC}, respectively.

\subsection{Definitions Reproduced}
\label{sec:rep}

Recall that $(\Xol,\Wol,S) \sim p(\xol,\wol,s)$ and
$\{(\Xol(k),\Wol(k),S(k))\}$ are drawn i.i.d. $\sim p(\xol,\wol,s)$. The subset $\Xol_J$, $J\subseteq I_M$, of sources are losslessly decoded and 
the complementary subset $\Xol_{J^c}$ is decoded  
under $L$ bounded distortion criteria $d_l:\Xcalol_{J^c}^{2} \rightarrow [0,d_{l\max}]$, $l\in I_L$. 
Any $(\Rol,\Dol)\in \Acal$ if for any $\e>0$ there exists encoder mappings 
$$
\begin{array}{rcll} 
f_{m}&:&\Xcal_m^n \rightarrow \Zcal_m, \quad &m\in I_M\\
f_{M+m}&:&\Wcal_m^n \rightarrow \Zcal_{M+m}, \quad &m\in I_K
\end{array}
$$
(i.e., $\fol:\Xcalol^{n}\times \Wcalol^n  \rightarrowtail \Zcalol$) and decoder mapping
$g: \Zcalol\times\Scal^n \rightarrow \Xcalol^n$ such that (reproducing (\ref{eq:AR'1}), (\ref{eq:ASS1}) and (\ref{eq:ASS2}), respectively)
\bea
\label{eq:AA1}
\frac{1}{n} \log |\Zcal_m | &\le& R_m +\e,\quad m\in I_{M+K}\\
\label{eq:AA2}
\Pr\{\Xol_J^n\ne \Xolhat_J^n)\} &\le& \e\\
\label{eq:AA3}
\frac{1}{n} \E d_{ln}(\Xol_{J^c}^{n}, \Xolhat_{J^c}^{n}) &\le&  D_l + \e,
\quad l\in I_L
\eea
where $\Xolhat^n =g( \fol(\Xol^n,\Wol^n),S^n)$. 

Further,
any $(\Rol,\Dol)\in\Acal_n^{*}$ if
there exist product of $(M+K)$ alphabets $\Zcalol$ (with the restriction $\Zcal_m=\Xcal_m^n$, $m\in J$),
conditional distributions $q_{m}(z_m|x_m^n)$, $m\in J^c$, and $r_{j}(z_{M+j}|w_j^n)$, $j\in I_K$, and mapping $\psi:\Zcalol \rightarrow\Xcalol^n_{J^c}$ such that (reproducing (\ref{eq:R1I.})--(\ref{eq:D.}), respectively)
\bea
\label{eq:AA1*}
\onen
I(\Xol_I^n;\Zol_I|\Zol_{I_{M+K}\setminus I},S^n) &\le& R_I, \quad I\subseteq I_M\setminus \{\}\\
\label{eq:AA2*}
\onen I(\Wol_I^n;\Zol_{M+I}|\Zol_{M+ I_K\setminus I},S^n) &\le& R_{M+I}, \quad I\subseteq I_K\setminus \{\}\\
\label{eq:AA3*}
\onen d_{ln}(\Xol^n_{J^c},\psi(\Zol,S^n))&\le& D_l,\quad l\in I_L
\eea
where
\bea
\label{eq:AA4*}
\Zol_J &=& \Xol_J^n\\
\label{eq:AA5*}
(\Xol^n,\Wol^n,S^n,\Zol_{I_{M+K}\setminus J}) &\sim& p_n(\xol^n,\wol^n,s^n)
\prod_{m\in J^c} q_{m}(z_m|x_m^n)
\prod_{j\in I_K} r_{j}(z_{M+j}|w_j^n)
\eea
(reproducing (\ref{eq:pTPC}))
and $p_n(\xol^n,\wol^n,s^n) = \prod_{k=1}^n p(\xol(k),\wol(k),s(k))$. Moreover, we have seen that, splitting $I={I'\cup I''}\subseteq I_M\setminus \{\}$ such that $I'\subseteq J$ and $I''\subseteq J^c$, we can equivalently write (\ref{eq:AA1*}) as (reproducing (\ref{eq:TPC.}))
\be
\label{eq:AA6*}
\onen H(\Xol_{I'}^n|\Xol^n_{J\setminus I'}, 
\Zol_{I_{M+K}\setminus(J\cup I'')},S^n)
+
\onen I(\Xol_{I''}^n;\Zol_{I''}|\Xol^n_{J}, 
\Zol_{I_{M+K}\setminus(J\cup I'')},S^n)\le R_{I'}+R_{I''}.
\ee
Note that condition (\ref{eq:AA4*}) affects neither (\ref{eq:AA6*}) nor (\ref{eq:AA2*}). Hence, from the set of conditions defining $\Acal^*_n$, we remove (\ref{eq:AA4*}) and the requirement $\Zcalol_J=\Xcalol_J^n$ by absorbing it in (\ref{eq:AA3*}) so that
\bea
\label{eq:AA7*}
\onen d_{ln}(\Xol^n_{J^c},\psi(\Xol^n_{J},\Zol_{I_{M+K}\setminus J},S^n))&\le& D_l,\quad l\in I_L.
\eea
Finally, recall $\Acal^* = \cup_{n=1}^\infty \Acal_n^*$.

\subsection{Inner Bound \boldmath{$\Acal \supseteq \overline{\Acal^*}$}} 
\label{sec:2In}

For any $(\Rol,\Dol)\in \Acal^* = \bigcup_{n=1}^\infty \Acal_n^*$,
 (\ref{eq:AA1*})--(\ref{eq:AA5*}) hold for some $n$, $\{q_m\}$, $\{r_j\}$ and $\psi$. Now, referring to  Theorem \ref{th:funda}, identify $(\Yol,\Zol,V)$ with $(\Wol^n,\Zol_{M+I_K},S^n)$ and $\Rol'$ with $n\Rol_{M+I_K}$, and 
note that condition (\ref{eq:AA2*}) is same as (\ref{eq:Rdef}). 
Consequently, by Theorem \ref{th:funda}, for any $\e' \rightarrow 0$, there exists a sequence of mapping pairs
$$(\fol_{M+I_K}:\Wcalol^{nn'} \rightarrowtail \Ucalol_{M+I_K},g':\Ucalol_{M+I_K}\times\Scal^{nn'} \rightarrow \Zcalol_{M+I_K}^{n'})$$ 
(for some $n'\rightarrow \infty$)
such that (\ref{eq:rate}) and (\ref{eq:prob}) hold. In other words, we respectively have (the first condition (\ref{eq:rate}) is divided throughout by $n$)
\bea
\label{eq:rateJ}
\frac{1}{nn'} \log |\Ucal_m| &\le& R_m + \e'_1/n, \quad m\in M+I_K\\
\label{eq:probJ}
\Pr\{\Ecal\} &\le& \e'_1
\eea
where 
\beann
{\cal E} &=& \{ (\Wol^{nn'},\Zolhat_{M+I_K}^{n'},S^{nn'}) \notin \Tenn(\Wol^n,\Zol_{M+I_K},S^n)\}\\
\Zolhat_{M+I_K}^{n'} &=& g'(\fol_{M+I_K}(\Wol^{nn'}),S^{nn'})
\eeann 
and $\e'_1\rightarrow 0$.

Referring to  Theorem \ref{th:funda} again, now identify $(\Yol,\Zol,V)$ with $(\Xol^n,\Zol_{I_M},(\Zol_{M+I_K},S^n))$ and $\Rol'$ with $n\Rol_{I_M}$, and 
note that condition (\ref{eq:AA1*}) is same as (\ref{eq:Rdef}). Further, noting $\Xol^n \rightarrow (\Wol^n,S^n) \rightarrow \Zol_{M+I_K}$ forms 
Markov chain and applying Lemma~\ref{le:Mar} in view of (\ref{eq:probJ}), we obtain 
$$
\Pr\{(\Xol^{nn'},\Wol^{nn'},\Zolhat_{M+I_K}^{n'},S^{nn'}) \notin \Tenn(\Xol^n,\Wol^n,\Zol_{M+I_K},S^n)\} \le \e'_2
$$
where $\e'_2\rightarrow 0$ as $\e'\rightarrow 0$. Hence, of course,
$$
\Pr\{(\Xol^{nn'},\Zolhat_{M+I_K}^{n'},S^{nn'}) \notin \Tenn(\Xol^n,\Zol_{M+I_K},S^n)\} \le \e'_2.
$$
Consequently, by Theorem \ref{th:funda}, for the same $\e'$ as earelier, there exists a sequence of mapping pairs
$$(\fol_{I_M}:\Xcalol^{nn'} \rightarrowtail \Ucalol_{I_M},g'':\Ucalol_{I_M}\times(\Zcalol_{M+I_K}^{n'}\times\Scal^{nn'}) \rightarrow \Zcalol_{I_M}^{n'})$$ (for some $n'\rightarrow \infty$)
such that (\ref{eq:rate}) and (\ref{eq:prob}) hold. Specifically, we respectively have (the first condition (\ref{eq:rate}) is divided throughout by $n$)
\bea
\label{eq:rateJ'}
\frac{1}{nn'} \log |\Ucal_m| &\le& R_m + \e'_3/n, \quad m\in I_M\\
\label{eq:probJ'}
\Pr\{\Ecal'\} &\le& \e'_3
\eea
where 
\beann
{\cal E'} &=& \{ (\Xol^{nn'},\Zolhat^{n'}_{I_M},(\Zolhat_{M+I_K}^{n'},S^{nn'})) \notin \Tenn(\Xol^n,\Zol_{I_M},(\Zol_{M+I_K},S^n))\}\\
\Zolhat_{I_M}^{n'} &=& g''(\fol_{I_M}(\Xol^{nn'}),(\Zolhat_{M+I_K}^{n'},S^{nn'}))
\eeann 
and $\e'_3\rightarrow 0$. 

Recall $\Zol_J = \Xol_J^n$ (as mentioned in (\ref{eq:AA4*})) and write $\Xolhat_J^{nn'}=\Zolhat_J^{n'}$ 
so that,
from (\ref{eq:probJ'}), we immediately have 
\be
\label{eq:nop}
\Pr\{\Xol_J^{nn'}\ne \Xolhat_J^{nn'}\}\le \e'_3.
\ee
Further, denote
\bea
\label{eq:dui1}
\Xolhat_{J^c}^{n}(j) &=& \psi(\Zolhat(j),S^n(j)),\quad 
j\in I_{n'}.
\eea
Hence, from (\ref{eq:probJ'}), we have 
\be
\label{eq:dui2}
\Pr\{\Ecal_1\}
=
\Pr\{ (\Xol_{J^c}^{nn'},\Xolhat_{J^c}^{nn'}) \notin \Tenn(\Xol_{J^c}^n,\psi(\Zol,S^n))\} \le \e'_3.
\ee
Now note
\be
\label{eq:dui3}
\frac{1}{nn'}d_{l(nn')}(\xol_{J^c}^{nn'},\xolhat_{J^c}^{nn'})\le D_l +\e'\dlmax, \quad l\in I_L
\ee
for any $(\xol_{J^c}^{nn'},\xolhat_{J^c}^{nn'})\in \Tenn(\Xol_{J^c}^{n},\psi(\Zol,S^n))$. Hence, 
we obtain (for each $l\in I_L$)
\bea
%\label{eq:xc2}
\nonumber
\frac{1}{nn'} 
\E \,d_{l(nn')}(\Xol^{nn'}_{J^c},\Xolhat^{nn'}_{J^c})
&\le& (1- \Pr\{ {\cal E}_1\}) (D_l +\e'\dlmax) +  \Pr\{ {\cal E}_1\} \dlmax\\
\label{eq:xc3}
&\le& D_l +(\e' + \e'_3) \dlmax
\eea
 due to (\ref{eq:dui2}). 
 
 At this point, observe that, stacking $\Xolhat_{J^c}^{n}(j)$, $j\in I_{n'}$, in (\ref{eq:dui1}), we can write 
\be
\label{eq:xc4}
\Xolhat_{J^c}^{nn'} =
\psi'(\Zolhat^{n'},S^{nn'})
\ee
for certain mapping $\psi'$.
Hence, write
\beann
\Xolhat^{nn'} &=& (\Xolhat_{J}^{nn'},\Xolhat_{J^c}^{nn'}) \\ 
&=&(\Zolhat_J^{n'},\psi'((\Zolhat_{I_M}^{n'},\Zolhat_{M+I_K}^{n'}),S^{nn'}))
\eeann 
and recall $J \subseteq I_M$ as well as
\beann
\Zolhat_{I_M}^{n'} &=& g''(\fol_{I_M}(\Xol^{nn'}),(\Zolhat_{M+I_K}^{n'},S^{nn'}))\\ \Zolhat_{M+I_K}^{n'} &=& g'(\fol_{M+I_K}(\Wol^{nn'}),S^{nn'})
\eeann
to conclude that $\Xolhat^{nn'} = g(\fol(\Xol^{nn'},\Wol^{nn'}),S^{nn'})$ for certain $g: \Ucalol\times\Scal^{nn'} \rightarrow \Xcalol^{nn'}$. Moreover, for any $\e>0$,
choose
$\e'>0$ such that $\max\{\e'_1,\e'_3,(\e' + \e'_3) d_{\max}\}\le \e$. Consequently, for mapping pair $(\fol,g)$, conditions (\ref{eq:rateJ}) and (\ref{eq:rateJ'}) give rise to (\ref{eq:AA1}), (\ref{eq:nop}) to (\ref{eq:AA2}) and (\ref{eq:xc3}) to (\ref{eq:AA3}), with $(\Ucalol,nn')$ now playing the role of $(\Zcalol,n)$. Hence $(\Rol,\Dol)\in \Acal$. In other words, $\Acal\supseteq\Acal^*$. Since $\Acal$ is closed, we have $\Acal\supseteq\AcalSol$ (noting $\AcalSol$ is the smallest closed set with $\Acal^*$ as a subset).
This completes the proof.
\hfill$\Box$

\subsection{Outer Bound \boldmath{$\Acal \subseteq \overline{\Acal^*}$}} 
\label{sec:2Out}

The proof requires Fano's inequality, a weakened version of which states the following: Given random variables $U$ and $V$,
\be
\label{eq:Fano}
H(U|V)\le 1+\log|\Ucal|\Pr\{U\ne g(V)\}
\ee  
for any $g:\Vcal\rightarrow\Ucal$  \cite{Cover}. 

Now consider any $(\Rol,\Dol)\in \Acal$. Then, for any $\e>0$, by definition, there exists mapping pair
$$(\fol:\Xcalol^{n}\times \Wcalol^n  \rightarrowtail \Zcalol,g: \Zcalol\times\Scal^n \rightarrow \Xcalol^n)$$ of some length $n$ such that (\ref{eq:AA1})--(\ref{eq:AA3}) hold. 
We can further encode $\Zol_{M+I_K}=\fol_{M+I_K}(\Wol^{n})$ in a noncooperative manner 
with complete side information $S^n$ using interposed Slepian-Wolf code 
$$(\fol'_{M+I_K}:\Zcalol_{M+I_K}^{n'} \rightarrowtail \Ucalol_{M+I_K},g'_{2}:\Ucalol_{M+I_K}\times\Scal^{nn'} \rightarrow \Zcalol_{M+I_K}^{n'}).$$
Given $(\Rol,\Dol)\in \Acal$, $\e$ and $(\fol,g)$, 
any rate-vector $\Rol'_{M+I_K}$ is said to be achieved using interposed codes of the form $(\fol'_{M+I_K},g'_{2})$ if for any $\e'>0$ there exists such code (of length $n'$) that satisfies
\bea
\label{eq:ei1}
\frac{1}{n'} \log |\Ucal_m| &\le& R'_m +\e', \quad m\in M+I_K \\
\label{eq:ei2}
\Pr\{\Zol_{M+I_K}^{n'}\ne g'_2(\fol'_{M+I_K}(\Zol_{M+I_K}^{n'}),S^{nn'})\} &\le & \e'.
\eea
In view of (\ref{eq:AA1}),
setting $\fol'_{M+I_K}$ to identity mapping (clearly, $n'=1$, $\Ucalol_{M+I_K}=\Zcalol_{M+I_K}$) and choosing $g'_2(\fol'_{M+I_K}(\Zol_{M+I_K}^{n'}),S^{nn'})=\fol'_{M+I_K}(\Zol_{M+I_K}^{n'}) = \Zol_{M+I_K}^{n'}$, of course, (\ref{eq:ei1}) and (\ref{eq:ei2}) trivially hold for $\Rol'_{M+I_K} = n(\Rol_{M+I_K}+\e)$ irrespective of $\e'$. Therefore,
we have
\be
\label{eq:R1.'}
\onen H(\fol_{M+I}(\Wol_I^n)|\fol_{M+ I_K\setminus I}(\Wol_{I_K\setminus I}^n),S^n) \le \sum_{i\in I} (R_{M+i} +\e)
\le R_{M+I} +K\e, \quad I\subseteq I_K\setminus \{\}
\ee
by Slepian-Wolf theorem \cite{Cover}. 

Similarly, we can also encode $\Zol_{I_M}=\fol_{I_M}(\Xol^{n})$ with complete side information $(\Zol_{M+I_K},S^n)$ 
(recall $\Zol_{M+I_K} = \fol_{M+I_K}(\Wol^n)$)
using interposed Slepian-Wolf code 
$$(\fol'_{I_M}:\Zcalol_{I_M}^{n'} \rightarrowtail \Ucalol_{I_M},g'_{1}:\Ucalol_{I_M}\times(\Zcalol_{M+I_K}^{n'}\times\Scal^{nn'}) \rightarrow \Zcalol_{I_M}^{n'}).$$
Again,
given $(\Rol,\Dol)\in \Acal$, $\e$ and $(\fol,g)$, 
any rate-vector $\Rol'_{I_M}$ is said to be achieved using interposed codes of the form $(\fol'_{I_M},g'_{1})$ if for any $\e'>0$ there exists such code (of length $n'$) that satisfies
\bea
\label{eq:ei1'}
\frac{1}{n'} \log |\Ucal_m| &\le& R'_m +\e', \quad m\in I_M \\
\label{eq:ei2'}
\Pr\{\Zol_{I_M}^{n'}\ne g'_1(\fol'_{I_M}(\Zol_{I_M}^{n'}),
(\Zol_{M+I_K}^{n'},S^{nn'}))\} &\le & \e'.
\eea
In view of (\ref{eq:AA1}),
setting $\fol'_{I_M}$ to identity mapping (clearly, $n'=1$, $\Ucalol_{I_M}=\Zcalol_{I_M}$) and choosing $g'_1(\fol'_{I_M}(\Zol_{I_M}^{n'}),
(\Zol_{M+I_K}^{n'},S^{nn'}))=\fol'_{I_M}(\Zol_{I_M}^{n'}) = \Zol_{I_M}^{n'}$, of course, (\ref{eq:ei1'}) and (\ref{eq:ei2'}) trivially hold for $\Rol'_{I_M} = n(\Rol_{I_M}+\e)$ irrespective of $\e'$. 
Therefore,
we have
\be
\label{eq:R1.''}
\onen H(\fol_{I}(\Xol_I^n)| \fol_{I_M\setminus I}(\Xol_{I_M\setminus I}^n),
\fol_{M+ I_K}(\Wol^n),S^n) \le \sum_{i\in I} (R_{i} +\e)
\le R_{I} +M\e, \quad I\subseteq I_M\setminus \{\}
\ee
by Slepian-Wolf theorem \cite{Cover}. 

At this point, let us write $g=(g_{J},g_{J^c})$ such that the ranges of $g_{J}$ and $g_{J^c}$ are contained in $\Xcalol_J^n$ and $\Xcalol_{J^c}^n$, respectively.
Now, noting $\Xolhat_J^n=g_J(\fol(\Xol^n,\Wol^n),S^n))$ and applying Fano's inequality (\ref{eq:Fano}) for $U=\Xol_J^n$ and $V=(\fol(\Xol^n,\Wol^n),S^n)$,
we have
\be
\label{eq:Fano1}
H(\Xol_J^n|\fol(\Xol,\Wol),S^n) \le 1 + n\log|\Xcalol_{J}|\Pr\{\Xol_J^n\ne \Xolhat_J^n)\} \le n(1+\log|\Xcalol_{J}|)\e.
\ee
In the above, the second inequality follows by (\ref{eq:AA2}) and by choosing $n>1/\e$. Next, refer to (\ref{eq:R1.''}) and split each $I={I'\cup I''}\subseteq I_M\setminus \{\}$ such that $I'\subseteq J$ and $I''\subseteq J^c$ so that
\bea
\nonumber
R_{I'}+R_{I''} +M\e
&\ge&
\onen H(\fol_{I'}(\Xol_{I'}^n),\fol_{I''}(\Xol_{I''}^n)
| \fol_{I_M\setminus I}(\Xol_{I_M\setminus I}^n),
\fol_{M+ I_K}(\Wol^n),S^n)\\
\nonumber
&=&
\onen H(\Xol_{I'}^n,\fol_{I'}(\Xol_{I'}^n),\fol_{I''}(\Xol_{I''}^n)
| \fol_{I_M\setminus I}(\Xol_{I_M\setminus I}^n),
\fol_{M+ I_K}(\Wol^n),S^n)\\
\label{eq:cham}
&& \qquad 
- ~\onen H(\Xol_{I'}^n
| \fol_{I_M}(\Xol_{I_M}^n),
\fol_{M+ I_K}(\Wol^n),S^n)
\eea
due to the chain rule of entropy.
From (\ref{eq:Fano1}), note that
\be
\label{eq:Fano2}
H(\Xol_{I'}^n|\fol_{I_M}(\Xol_{I_M}^n),
\fol_{M+ I_K}(\Wol^n),S^n) \le n(1+\log|\Xcalol_{J}|)\e
\ee
(recall $\fol(\Xol,\Wol)=(\fol_{I_M}(\Xol_{I_M}^n),
\fol_{M+ I_K}(\Wol^n))$).
Further, we can write
\bea
\nonumber
&&
\!\!\!\!\!\!\!\!
%\!\!\!\!
%\!\!\!\!
H(\Xol_{I'}^n,\fol_{I'}(\Xol_{I'}^n),\fol_{I''}(\Xol_{I''}^n)
| \fol_{I_M\setminus I}(\Xol_{I_M\setminus I}^n),
\fol_{M+ I_K}(\Wol^n),S^n)\\
\nonumber
&&
=~~ 
H(\Xol_{I'}^n,\fol_{I''}(\Xol_{I''}^n)
| \fol_{I_M\setminus I}(\Xol_{I_M\setminus I}^n),
\fol_{M+ I_K}(\Wol^n),S^n)\\
\label{eq:cham'}
&& 
=~~
H(\Xol_{I'}^n,\fol_{I''}(\Xol_{I''}^n)
| \fol_{J\setminus {I'}}(\Xol_{J\setminus {I'}}^n),
\fol_{I_M\setminus (J\cup I'')}(\Xol_{I_M\setminus (J\cup I'')}^n),
\fol_{M+ I_K}(\Wol^n),S^n)\quad\\
\label{eq:cham''}
&& 
\ge~~
H(\Xol_{I'}^n,\fol_{I''}(\Xol_{I''}^n)
| \Xol_{J\setminus {I'}}^n,
\fol_{I_M\setminus (J\cup I'')}(\Xol_{I_M\setminus (J\cup I'')}^n),
\fol_{M+ I_K}(\Wol^n),S^n)\\
\nonumber
&&
=~~
H(\Xol_{I'}^n
| \Xol_{J\setminus {I'}}^n,
\fol_{I_M\setminus (J\cup I'')}(\Xol_{I_M\setminus (J\cup I'')}^n),
\fol_{M+ I_K}(\Wol^n),S^n) \\
\label{eq:cham'''}
&&\qquad
+~
H(\fol_{I''}(\Xol_{I''}^n)
| \Xol_{J}^n,
\fol_{I_M\setminus (J\cup I'')}(\Xol_{I_M\setminus (J\cup I'')}^n),
\fol_{M+ I_K}(\Wol^n),S^n).
\eea
Here (\ref{eq:cham'}) follows by noting
$I_M\setminus I = (J\setminus {I'})\cup (I_M\setminus(J\cup{I''}))$, 
(\ref{eq:cham''}) follows due to data processing inequality and (\ref{eq:cham'''}) follows by the chain rule of entropy. Using (\ref{eq:cham'''}) and (\ref{eq:Fano2}) in (\ref{eq:cham}) and rearranging, we obtain
\bea
\nonumber
&&
\!\!\!\!\!\!\!\!\!\!\!\!\!\!\!\!
\!\!\!\!\!\!
\onen 
H(\Xol_{I'}^n
| \Xol_{J\setminus {I'}}^n,
\fol_{I_M\setminus (J\cup I'')}(\Xol_{I_M\setminus (J\cup I'')}^n),
\fol_{M+ I_K}(\Wol^n),S^n)\\
\nonumber
&&
\!\!\!\!\!\!\!\!\!\!\!\!\!\!\!\!
+~
\onen 
H(\fol_{I''}(\Xol_{I''}^n)
| \Xol_{J}^n,
\fol_{I_M\setminus (J\cup I'')}(\Xol_{I_M\setminus (J\cup I'')}^n),
\fol_{M+ I_K}(\Wol^n),S^n)\\
\label{eq:cha}
&&
\le~ R_{I'} + R_{I''} +(M+1+\log|\Xcalol_{J}|)\e,\quad I'\subseteq J,I''\subseteq J^c,{I'\cup I''}\subseteq I_M\setminus \{\}.
\eea

Further, from (\ref{eq:AA3}), we have, for each $l\in I_L$,
\bea
\nonumber
D_l+\e
&\ge&
\frac{1}{n} \E d_{ln}(\Xol_{J^c}^{n}, g_{J^c}(\fol_{I_M}(\Xol_{I_M}^n),
\fol_{M+ I_K}(\Wol^n),S^n)) \\
\label{eq:g2'}
&=&
\onen 
\E d_{ln}(\Xol_{J^c}^{n}, g'_{J^c}(\Xolhat^n_{J},
\fol_{I_M}(\Xol_{I_M}^n),
\fol_{M+ I_K}(\Wol^n),S^n))\\
\nonumber
&\ge& (1-\Pr\{\Xol_J^{n}\ne \Xolhat_J^{n}\}) \onen 
\E d_{ln}(\Xol_{J^c}^{n}, g'_{J^c}(\Xol^n_{J},
\fol_{I_M}(\Xol_{I_M}^n),
\fol_{M+ I_K}(\Wol^n),S^n))\\
\label{eq:edmax}
&\ge& \onen 
\E d_{ln}(\Xol_{J^c}^{n}, g'_{J^c}(\Xol^n_{J},
\fol_{I_M}(\Xol_{I_M}^n),
\fol_{M+ I_K}(\Wol^n),S^n)) -\e\dlmax\\
\nonumber
&\ge& \onen 
\E d_{ln}(\Xol_{J^c}^{n}, g'_{J^c}(\Xol^n_{J},(\fol_J(\Xol^n_{J}),
\fol_{I_M\setminus J}(\Xol_{I_M\setminus J}^n)),
\fol_{M+ I_K}(\Wol^n),S^n)) -\e\dlmax\\
\label{eq:psix}
&=& \onen 
\E d_{ln}(\Xol_{J^c}^{n}, \psi(\Xol^n_{J},
(\fol_{I_M\setminus J}(\Xol_{I_M\setminus J}^n),
\fol_{M+ I_K}(\Wol^n)),S^n)) -\e\dlmax
\eea
where (\ref{eq:g2'}) clearly holds for suitable $g_{J^c}'$, (\ref{eq:edmax}) follows because $\Pr\{\Xol_J^{n}\ne \Xolhat_J^{n}\}\le \e$ (by (\ref{eq:AA2})) and (\ref{eq:psix}) clearly holds for suitable $\psi$. Rearranging (\ref{eq:psix}), we obtain
\be
\label{eq:psix1}
\onen 
\E d_{ln}(\Xol_{J^c}^{n}, \psi(\Xol^n_{J},
(\fol_{I_M\setminus J}(\Xol_{I_M\setminus J}^n),
\fol_{M+ I_K}(\Wol^n)),S^n)) \le D_l +(1+\dlmax)\e,\quad l\in I_L. 
\ee

At this point define for any $\e\ge 0$ and any integral $n\ge 1$ the set $\Acal_n^{*(\e)}$ of rate-distortion pairs $(\Rol,\Dol)$ such that there exist product $\Zcalol$ of $M+K$ alphabets (of which $\Zcalol_J$ can be arbitrarily chosen), 
conditional distributions $q_{m}(z_m|x_m^n)$, $m\in J^c$, and $r_{j}(z_{M+j}|w_j^n)$, $j\in I_K$, and mapping $\psi:\Xcalol_J^n\times\Zcalol_{I_{M+K}\setminus J} \rightarrow\Xcalol^n_{J^c}$ such that 
\bea
\nonumber
&&
\!\!\!\!\!\!\!\!\!\!\!\!\!\!\!\!\!\!\!\!\!\!\!\!\!\!\!\!\!\!\!\!\!\!\!\!\!\!\!\!
\!\!\!\!\!\!\!\!\!\!\!\!\!\!\!\!\!\!\!\!\!\!\!\!\!\!\!\!\!\!\!\!\!\!\!\!\!\!\!\!
%\!\!\!\!\!\!\!\!\!\!\!\!\!\!\!\!\!\!\!\!
\!\!\!\!\!\!
\!\!\!\!\!\!\!\!\!\!\!\!\!\!
\onen H(\Xol_{I'}^n|\Xol^n_{J\setminus I'}, 
\Zol_{I_{M+K}\setminus(J\cup I'')},S^n)
%\nonumber
+
\onen I(\Xol_{I''}^n;\Zol_{I''}|\Xol^n_{J}, 
\Zol_{I_{M+K}\setminus(J\cup I'')},S^n)
%\qquad
\\
\nonumber
&\le& R_{I'}+R_{I''}
+(M+1+\log|\Xcalol_{J}|)\e,\\
\label{eq:AA6*'}
&&
\quad I'\subseteq J,I''\subseteq J^c,{I'\cup I''}\subseteq I_M\setminus \{\}
\\
\label{eq:AA2*'}
\onen I(\Wol_I^n;\Zol_{M+I}|\Zol_{M+ I_K\setminus I},S^n) &\le& R_{M+I}
+K\e, \quad I\subseteq I_K\setminus \{\}\\
\label{eq:AA7*'}
\onen d_{ln}(\Xol^n_{J^c},\psi(\Xol^n_{J},\Zol_{I_{M+K}\setminus J},S^n))&\le& D_l+(1+\dlmax)\e,\quad l\in I_L
\eea
where
\bea
\label{eq:AA5*'}
(\Xol^n,\Wol^n,S^n,\Zol_{I_{M+K}\setminus J}) &\sim& p_n(\xol^n,\wol^n,s^n)
\prod_{m\in J^c} q_{m}(z_m|x_m^n)
\prod_{j\in I_K} r_{j}(z_{M+j}|w_j^n).
\eea
Comparing (\ref{eq:AA6*'})--(\ref{eq:AA5*'}) with (\ref{eq:AA6*}), (\ref{eq:AA2*}), (\ref{eq:AA7*}) and (\ref{eq:AA5*}), respectively, note that $\Acal_n^{*(0)}=\Acal_n^*$.
Also note that
$\Acal_n^{*(\e_1)}\subseteq\Acal_n^{*(\e_2)}$ for $\e_1\le \e_2$.
Further, let $\Acal^{*(\e)} = \bigcup_{n=1}^\infty \Acal_n^{*(\e)}$.
Of course, $\Acal^{*(\e_1)}\subseteq\Acal^{*(\e_2)}$ for $\e_1\le \e_2$. 
Hence, noting $\bigcap_{\e>0} \overline{\Acal^{*(\e)}}$ is closed, we obtain
\be
\label{eq:lope}
\bigcap_{\e>0} \overline{\Acal^{*(\e)}} = \overline{\Acal^{*(0)}} = \overline{\Acal^*}.
\ee
The second equality in (\ref{eq:lope}) holds because $\Acal^{*(0)} = \bigcup_{n=1}^\infty \Acal_n^{*(0)}= \bigcup_{n=1}^\infty \Acal_n^*= \Acal^*$.

Finally, consider any $(\Rol,\Dol)\in \Acal$. 
Recall that for any $\e>0$ there exists mapping pair $(\fol,g)$ such that (\ref{eq:cha}), (\ref{eq:R1.'}), and (\ref{eq:psix1}) hold. Choosing 
$$\Zol_{I_{M+K}\setminus J} = (\fol_{J^c}(\Xol_{J^c}^n),\fol_{M+I_K}(\Wol^n))$$
and keeping the present $\psi$, note that the above three conditions coincide with (\ref{eq:AA6*'})-(\ref{eq:AA7*'}), respectively. Also, (\ref{eq:AA5*'}) holds for $q_m(z_m|x_m^n) =\delta(z_m-x_m^n)$, $m\in J^c$, and $r_j(z_{M+j}|w_j^n) =\delta(z_{M+j}-w_j^n)$, $j\in I_K$, where $\delta(a-b)=1$ if $a=b$ and $\delta(a-b)=0$ otherwise. 
Hence $(\Rol,\Dol)\in \Acal_n^{*(\e)}$. 
Consequently, we have $(\Rol,\Dol)\in \Acal^{*(\e)}\subseteq \overline{\Acal^{*(\e)}}$ for each $\e>0$. Hence, by (\ref{eq:lope}), we have $(\Rol,\Dol)\in \bigcap_{\e>0} \overline{\Acal^{*(\e)}}=\overline{\Acal^*}$.
This completes the proof. \hfill$\Box$

\Section{Application to Estimation Theory}
\label{sec:decision}
Next we pose the entropy-constrained estimation problem in a multiterminal setting. We shall show that our theory of source coding solves this problem as a special case.

{\bf {\em Problem Statement}:} Consider estimation of $X$ on the basis of  observations $\Wol = (W_1,W_2,...,W_K)$ available at base station at respective rates $\Rol = (R_1,R_2,...,R_K)$. In addition, let observation $S$ be completely available at the base station. 
The estimation error is measured using a bounded distortion criterion 
$d:\Xcalol^{2} \rightarrow [0,d_{\max}]$.
Formally, 
suppose $(X,\Wol,S) \sim p(x,\wol,s)$ and
draw $\{(X(k),\Wol(k),S(k))\}$ i.i.d. $\sim p(x,\wol,s)$. We encode $\Wol$ using $K$ encoder mappings
\be
\label{eq:fm'}
f_{m}:\Wcal_m^n \rightarrow \Zcal_m, \quad m\in I_K
\ee
(i.e., $\fol:\Wcalol^{n} \rightarrowtail \Zcalol$)
and decode using decoder mapping
\be
\label{eq:gC'}
g: \Zcalol\times\Scal^n \rightarrow \Xcal^n.
\ee
In other words, estimate $X^n$ by 
\be
\label{eq:rec''.}
\Xhat^n =g( \fol(\Wol^n),S^n)
\ee  
with corresponding estimation error $\onen \E d_n(X^n,\Xhat^n)$. The achievable set $\Ae$ 
is defined by the set of pairs $(\Rol',D)$ such that, 
for any $\e>0$, 
\bea
\label{eq:RE}
\onen \log |\Zcal_m|&\le& R'_m +\e,\quad m\in I_K\\
\label{eq:AE}
\frac{1}{n} \E d_{n}(X^{n}, \Xhat^{n}) &\le&  D + \e.
\eea
Next we give an information-theoretic description of $\Ae$.

{\bf {\em Achievable Region}:} First of all, refer to Sec.~\ref{sec:comp} and note that the estimation problem at hand is a special case of the source coding problem `DPC'. Specifically, the number of sources $\Xol$ is $M=1$ and the number of distortion criterion is $L=1$. Further, $X$ is not encoded, i.e., rate $R_1=0$, and $\Rol'$ now plays the role of $R_{1+I_K}$. 
Hence, by Theorem \ref{th:generic}, we have
$$\Ae =\overline{\Ae^*},$$
where $\Ae^*=\cup_{n=1}^\infty \AnE^*$,
\be
\label{eq:AnE}
\AnE^*=\{(\Rol_{1+I_K},D): (\Rol,D)\in \AnDPC^*, R_1=0\}
\ee
and $\AnDPC^*$ is defined for $M=1$ and $L=1$. Further, refer to the conditions (\ref{eq:'R1I})--(\ref{eq:pLCPC}) defining $\AnDPC^*$ and use $R_1=0$. Note that condition (\ref{eq:'R1I}) and 
the auxiliary random variable $Z_1$  
are redundant. Hence, $Z_1$ can be marginalized out from (\ref{eq:pLCPC}). For an explicit definition of $\AnE^*$ equivalent to (\ref{eq:AnE}), let us rechristen $\Zol_{1+I_K}$ as $\Zol'$ and rewrite (\ref{eq:'R2I})--(\ref{eq:pLCPC}) in the modified form. In particular, 
a rate-distortion pair $(\Rol',D)$ belongs to $\AnE^{*}$ if
there exist product of $K$ alphabets $\Zcalol'$,
conditional distributions $r_{j}(z'_{j}|w_j^n)$, $j\in I_K$, and mapping $\psi:\Zcalol' \rightarrow\Xcal^n$ such that
\bea
\label{eq:'.R2I}
\onen I(\Wol_I^n;\Zol'_{I}|\Zol'_{I^c},S^n) &\le& R'_{I}, \quad I\subseteq I_K\setminus \{\}\\
\label{eq:'.D}
\onen d_{n}(X^n,\psi(\Zol',S^n))&\le& D
\eea
where 
\be
\label{eq:.pLCPC}
(X^n,\Wol^n,S^n,\Zol') \sim p_n(x^n,\wol^n,s^n)
\prod_{j\in I_K} r_{j}(z'_{j}|w_j^n)
\ee 
and $p_n(x^n,\wol^n,s^n) = \prod_{k=1}^n p(x(k),\wol(k),s(k))$. This  solves the entropy-constraint estimation problem.

%\Section{Important Special Cases}
%\label{sec:impli}
%\input{impli}

\Section{Conclusion}
\label{sec:discuss}
In this paper, we presented a unified solution to all multiterminal source coding problems where encoders do not cooperate, encoded information is jointly decoded 
and the distortion criteria, if any, apply to single letters. 
In particular, we unify all admissible source coding problems, irrespective of number of sources and availability of side information, using a fundamental principle (Theorem \ref{th:funda}) based on typicality. 
The power of the above principle comes from 
a novel dissociation of distortion criteria from the core 
of source coding problems.
In a way, our work marks the culmination of decades of source coding research pioneered by Shannon \cite{ShanLL,Shannon} and enriched by works of Slepian-Wolf \cite{SW}, Wyner \cite{Wyner}, Ahlswede-K\"{o}rner \cite{AhlKor}, Wyner-Ziv \cite{WZ}, Berger {\em et al.} \cite{Upper} and Berger-Yeung \cite{BY}. At the same time,
our result clears
the path for new research which hitherto seemed too difficult to attempt. 
The multidecoder extension of our theory is of course the natural
next step. 
Another open problem that also comes to mind is characterization of the achievable region for the entropy-constrained detection problem in the multiterminal setting. The main difficulty in this problem is that natural performance measures of detection, such as Bayesian probability of error, are not of single-letter type. We also believe that the ongoing research into channel coding theory will receive certain direct and indirect clues from our work. In the least, researchers investigating the capacity regions of not-so-well-understood channels, such as the broadcast channel,
will now be open to the possibility of a higher order information-theoretic description instead of the usual first order.

\newpage
\appendix
\setcounter{equation}{0}
\renewcommand{\theequation}{\Alph{section}.\arabic{equation}}

\Section{Equivalence of (\ref{eq:R1I.}) and (\ref{eq:TPC.}) }
\label{app:TPC}
Recall $\Zol_J=\Xol_J^n$ and $I=I'\cup I''\subseteq I_M\setminus \{\}$ such that $I'\subseteq J$ and $I''\subseteq J^c$. Hence, we can write
\bea
\label{eq:ke0}
I(\Xol_I^n;\Zol_I|\Zol_{I_{M+K}\setminus I},S^n) 
&=& 
I(\Xol_{I'}^n,\Xol_{I''}^n ;\Xol_{I'}^n,\Zol_{I''}|\Xol^n_{J\setminus I'}, 
\Zol_{I_{M+K}\setminus(J\cup I'')},S^n)\\
\label{eq:ke1}
&=& I(\Xol_{I'}^n;\Xol_{I'}^n,\Zol_{I''}|\Xol^n_{J\setminus I'}, U) + 
I(\Xol_{I''}^n ;\Xol_{I'}^n,\Zol_{I''}|\Xol_{J}^n, U)\\
\label{eq:ke2}
&=& H(\Xol_{I'}^n|\Xol^n_{J\setminus I'}, U) + 
I(\Xol_{I''}^n ;\Zol_{I''}|\Xol_{J}^n, U)
\eea
where 
(\ref{eq:ke0}) follows by noting
$I_{M+K}\setminus I = (J\setminus {I'})\cup (I_{M+K}\setminus(J\cup{I''}))$,
(\ref{eq:ke1}) follows by the chain rule of mutual information 
and by denoting
$U= (\Zol_{I_{M+K}\setminus(J\cup I'')},S^n)$ and (\ref{eq:ke2}) follows because $I'\subseteq J$. In view of (\ref{eq:ke2}), (\ref{eq:R1I.}) and (\ref{eq:TPC.}) are indeed equivalent.

\Section{Proof of Lemma \ref{le:corner}}
\label{sec:cornerMom}
\subsection{Information-Theoretic Relations}
\label{sec:equal}

First we need certain information-theoretic relations involving $$(\Yol,\Zol,V)\sim p'(\yol,v)\prod_{m=1}^{M'} q'_{m}(u_m|y_m)$$
where all random variables are dependent.

\begin{lemma}
\label{le:chain'}
Suppose sets $I,I'\subseteq I_{M'}\setminus \{\}$ are disjoint. Then 
\be
\label{eq:rule'} 
I\left(\Yol_{I};\Zol_{I}|\Zol_{{(I\cup I')}^c},V\right)
= I\left(\Yol_{I};\Zol_{I}|\Zol_{{I}^c},V\right)+ I\left(\Zol_I;\Zol_{I'}|\Zol_{{(I\cup I')}^c},V\right).
\ee
\end{lemma}

{\bf {\em Proof}:} First expand
\be
\label{eq:ami1}
I\left(\Zol_I;\Yol_{I},\Zol_{I'}|\Zol_{{(I\cup I')}^c},V\right)
= I\left(\Zol_I;\Zol_{I'}|\Zol_{{(I\cup I')}^c},V\right) + I\left(\Zol_{I};\Yol_{I}|\Zol_{{I}^c},V\right),
\ee
applying the chain rule of mutual entropy. Expand the same quantity again, now applying the chain rule in a different order:
\be
\label{eq:ami2}
I\left(\Zol_I;\Yol_{I},\Zol_{I'}|\Zol_{{(I\cup I')}^c},V\right)
= I\left(\Zol_{I};\Yol_{I}|\Zol_{{(I\cup I')}^c},V\right)
+ I\left(\Zol_I;\Zol_{I'}|\Yol_{I},\Zol_{{(I\cup I')}^c},V\right).
\ee
Note that $\Zol_I\rightarrow (\Yol_{I},\Zol_{{(I\cup I')}^c},V) \rightarrow 
\Zol_{I'}$ forms Markov chain, i.e., 
$I\left(\Zol_I;\Zol_{I'}|\Yol_{I},\Zol_{{(I\cup I')}^c},V\right)
=0$ in (\ref{eq:ami2}). Hence, equating right hand sides of (\ref{eq:ami1})
and (\ref{eq:ami2})
and rearranging, we obtain (\ref{eq:rule'}). \hfill$\Box$

\begin{lemma}
\label{le:chain}
Suppose sets $I,I'\subseteq I_{M'}\setminus \{\}$ are disjoint. Then 
\bea
\label{eq:rule1}
I\left(\Yol_{I\cup I'};\Zol_{I\cup I'}|\Zol_{{(I\cup I')}^c},V\right)
&=& I\left(\Yol_{I};\Zol_{I}|\Zol_{{(I\cup I')}^c},V\right)
+ 
I\left(\Yol_{I'};\Zol_{I'}|\Zol_{{I'}^c},V\right).
\eea
\end{lemma}

{\bf {\em Proof}:} For any quadruple $(U_1,U_2;V_1,V_2)$ of random variables, we can write
\bea
\nonumber
I(U_1,U_2;V_1,V_2) &=& I(U_1,U_2;V_1) + I(U_1,U_2;V_2|V_1)\\
\label{eq:red1}
&=& 
I(U_1;V_1) + I(U_2;V_1|U_1) + I(U_2;V_2|V_1) + I(U_1;V_2|V_1,U_2)
\eea
by repeatedly applying the chain rule of mutual information. Using formula (\ref{eq:red1}), we obtain 
\bea
\nonumber
I\left(\Yol_{I\cup I'};\Zol_{I\cup I'}|\Zol_{{(I\cup I')}^c},V\right)
&=&
I\left(\Yol_{I};\Zol_{I}|\Zol_{{(I\cup I')}^c},V\right)
+
I\left(\Yol_{I'};\Zol_{I}|\Yol_I,\Zol_{{(I\cup I')}^c},V\right)\\
\label{eq:red2}
&&~~ +
I\left(\Yol_{I'};\Zol_{I'}|\Zol_{{I'}^c},V\right)
+
I\left(\Yol_{I};\Zol_{I'}|\Yol_{I'},\Zol_{{I'}^c},V\right).
\eea
Here 
$I\left(\Yol_{I'};\Zol_{I}|\Yol_I,\Zol_{{(I\cup I')}^c},V\right) =0$ and 
$I\left(\Yol_{I};\Zol_{I'}|\Yol_{I'},\Zol_{{I'}^c},V\right) =0$, respectively, because $\Zol_{I} \rightarrow (\Yol_I,\Zol_{{(I\cup I')}^c},V) \rightarrow \Yol_{I'}$ and $\Zol_{I'} \rightarrow (\Yol_{I'},\Zol_{{I'}^c},V) \rightarrow \Yol_{I}$ form Markov chains. Hence the result.
\hfill$\Box$

More generally, any $\Ihat\subseteq I_{M'}\setminus \{\}$ can play the role of $I_{M'}$ in the statement of Lemma \ref{le:chain} so that ${I'}^c$ can be replaced by $\Ihat\setminus I'$ and ${(I\cup I')}^c$ by $\Ihat \setminus ({I\cup I'})$. In that case, Lemma \ref{le:chain} immediately takes the form:

\begin{corollary}
\label{cor:chain1} Suppose
set $\Ihat \subseteq 
I_{M'}\setminus \{\}$ is given and
sets $I,I'\subseteq \Ihat$ are disjoint. Then 
\bea
\label{eq:rule1'}
I\left(\Yol_{I\cup I'};\Zol_{I\cup I'}|\Zol_{{\Ihat \setminus(I\cup I')}},V\right)
&=& I\left(\Yol_{I};\Zol_{I}|\Zol_{{\Ihat \setminus(I\cup I')}},V\right)
+ 
I\left(\Yol_{I'};\Zol_{I'}|\Zol_{{\Ihat \setminus I'}},V\right).
\eea
\end{corollary}

Now consider any $I\subseteq I_{M'}\setminus\{\}$ with cardinality $|I|=m$ and write $I=\{i(1;m)\}$. Further, setting $\Ihat=I_{M'}$ and letting $(\{i(1)\}, I\setminus\{i(1)\})$ play the role of $(I,I')$
in (\ref{eq:rule1'}), we have
\bea
\label{eq:eka1}
I\left(\Yol_{I};\Zol_{I}|\Zol_{I^c},V\right)
&=& I\left(Y_{i(1)};Z_{i(1)}|\Zol_{I^c},V\right)
+ 
I\left(\Yol_{I\setminus\{i(1)\}};\Zol_{I\setminus\{i(1)\}}|
\Zol_{{(I\setminus\{i(1)\})}^c},V\right).\quad
\eea
Noting $I\setminus\{i(1)\} = \{i(2:m)\}$ and
continuing the recursion by letting $(\{i(2)\}, I\setminus\{i(1;2)\})$ play the role of $(I,I')$
in (\ref{eq:rule1'}) and so on, we obtain
\bea
\label{eq:eka2}
I\left(\Yol_{I};\Zol_{I}|
\Zol_{I^c},V\right)
&=& \sum_{j=1}^m
I\left(Y_{i(j)};Z_{i(j)}|\Zol_{{(I\setminus \{i(1:j-1)\})}^c},V\right).
\eea
Noting ${(I\setminus \{i(1:j-1)\})}^c = I_{M'}\setminus \{i(j:m)\}$ in (\ref{eq:eka2}), we have the following:

\begin{corollary}
\label{cor:mid}
For any set $I=\{i(1;m)\}\subseteq I_{M'}\setminus\{\}$,
\bea
\label{eq:mid}
I\left(\Yol_{I};\Zol_{I}|\Zol_{I^c},V\right)
&=&
\sum_{j=1}^m
I\left(Y_{i(j)};Z_{i(j)}|\Zol_{I_{M'}\setminus \{i(j:m)\}},V\right).
\eea
\end{corollary}

Further,  suppose $\Ihat =I_m$ for some $2\le m\le M'$. For the choice $I= I_{m-1}$ and $I' = \{m\}$, (\ref{eq:rule1'}) becomes
\be
\label{eq:cc1}
I\left(\Yol_{I_m};\Zol_{I_m}|V\right)=
I\left(\Yol_{I_{m-1}};\Zol_{I_{m-1}}|V\right)
+I\left(Y_m;Z_m|\Zol_{I_{m-1}},V\right),
\ee
which gives a useful chain rule. Applying this repeatedly, we obtain:

\begin{corollary}
\label{cor:chain2}
For any $2\le m \le M'$,
\be
\label{eq:cc2}
I\left(\Yol_{I_{m}};\Zol_{I_{m}}|V\right)= 
\sum_{i=1}^{m} I\left(Y_i;Z_i|\Zol_{I_{i-1}},V\right).
\ee
\end{corollary}

In fact, corollary \ref{cor:chain2} can be further generalized as follows.
For any $1\le m < M'$, set $\Ihat = I_{M'}$, $I=I_m$ and $I'=I_{M'}\setminus I_m$ in Lemma \ref{cor:chain1} to obtain
\be
\label{eq:cc4}
I\left(\Yol_{I_{M'}};\Zol_{I_{M'}}|V\right) 
= I\left(\Yol_{I_{m}};\Zol_{I_{m}}|V\right)
+
I\left(\Yol_{I_{M'}\setminus I_m};\Zol_{I_{M'}\setminus I_m}|
\Zol_{I_{m}},V\right).
\ee
Expanding $I\left(\Yol_{I_{M'}};\Zol_{I_{M'}}|V\right)$ and 
$I\left(\Yol_{I_{m}};\Zol_{I_{m}}|V\right)$ using Corollary \ref{cor:chain2},
from (\ref{eq:cc4}) we obtain:

\begin{corollary}
\label{cor:chain3}
For any $1\le m < M'$,
\be
\label{eq:cc3}
I\left(\Yol_{I_{M'}\setminus I_m};\Zol_{I_{M'}\setminus I_m}|
\Zol_{I_{m}},V\right)= 
\sum_{i=m+1}^{M'} I\left(Y_i;Z_i|\Zol_{I_{i-1}},V\right).
\ee
\end{corollary}

We require one more information-theoretic relation.

\begin{lemma}
\label{le:last}
For any $I\subseteq I_{M'}\setminus \{\}$,
\be
\label{eq:last}
I\left(\Yol_{I};\Zol_{I}|\Zol_{I^c},V\right) \le
\sum_{i\in I} I\left(Y_i;Z_i|\Zol_{I_{i-1}},V\right).
\ee
\end{lemma}

{\bf {\em Proof}:} Let $m=|I|$ and write the elements of $I=\{i(1),i(2),...,i(m)\}$ in ascending order. Hence, we have
\be
\label{eq:last1}
I_{i(j)-1} \subseteq  I_{M'}\setminus \{i(j:m)\}, \quad j \in I_m.
\ee
For each $j$,
denote $$\Itil(j) = (I_{M'}\setminus \{i(j:m)\})\setminus I_{i(j)-1}.$$ 
Hence we can write
\bea
\nonumber
I\left(Y_{i(j)};Z_{i(j)}|\Zol_{I_{M'}\setminus \{i(j:m)\}},V\right)
&=& I\left(Y_{i(j)};Z_{i(j)}|\Zol_{I_{i(j)-1}}, \Zol_{\Itil(j)},V\right)\\
\nonumber
&=& H\left(Z_{i(j)}|\Zol_{I_{i(j)-1}}, \Zol_{\Itil(j)},V\right)
- 
H\left(Z_{i(j)}|Y_{i(j)},\Zol_{I_{i(j)-1}}, \Zol_{\Itil(j)},V\right)\\
\label{eq:last2}
&=& 
H\left(Z_{i(j)}|\Zol_{I_{i(j)-1}}, \Zol_{\Itil(j)},V\right)
- 
H\left(Z_{i(j)}|Y_{i(j)},\Zol_{I_{i(j)-1}}, V\right)\\
\label{eq:last3}
&\le&
H\left(Z_{i(j)}|\Zol_{I_{i(j)-1}}, V\right)
- 
H\left(Z_{i(j)}|Y_{i(j)},\Zol_{I_{i(j)-1}}, V\right)\\
\label{eq:last4}
&=& I\left(Y_{i(j)};Z_{i(j)}|\Zol_{I_{i(j)-1}}, V\right).
\eea
Here (\ref{eq:last2}) follows by noting
$$
H\left(Z_{i(j)}|Y_{i(j)},\Zol_{I_{i-1}}, \Zol_{\Itil(j)},V\right)
= H\left(Z_{i(j)}|Y_{i(j)}\right)
= H\left(Z_{i(j)}|Y_{i(j)},\Zol_{I_{i-1}},V\right)
$$
due to the fact that $Z_{i(j)} \rightarrow Y_{i(j)}
\rightarrow (\Zol_{I_{i-1}}, \Zol_{\Itil(j)},V)$ form Markov chain. Further, (\ref{eq:last3}) follows because conditioning reduces entropy. Now summing (\ref{eq:last4}) over $j\in I_m$, we obtain
\be
\label{eq:last5}
\sum_{j=1}^m I\left(Y_{i(j)};Z_{i(j)}|\Zol_{I_{M'}\setminus \{i(j:m)\}},V\right)
\le \sum_{j=1}^m I\left(Y_{i(j)};Z_{i(j)}|\Zol_{I_{i(j)-1}}, V\right).
\ee
By Corollary \ref{cor:mid}, the left hand side of (\ref{eq:last5}) equals $I\left(\Yol_{I};\Zol_{I}|\Zol_{I^c},V\right)$. Also, note that the right hand side of (\ref{eq:last5}) is same as the right hand side of (\ref{eq:last}). 
Hence (\ref{eq:last5}) is the desired result. \hfill $\Box$ 

Now let us turn our attention to finding the corner points of $\Bcal^*$.

\subsection{Properties of Active Constraints} 
\label{sec:constraint}

\begin{lemma}
\label{le:unique}
Suppose 
\bea
\label{eq:I}
I\left(\Yol_{I};\Zol_I|\Zol_{I^c},V\right) &=& R'_I\\
\label{eq:I'}
I\left(\Yol_{I'};\Zol_{I'}|\Zol_{{I'}^c},V\right) &=& R'_{I'}
\eea
simultaneously hold for distinct
sets $I,I'\subseteq I_M\setminus \{\}$. Then either $I\subset I'$ or $I'\subset I$.
\end{lemma}

{\bf {\em Proof}:} It is enough to show that if $I\setminus I'\ne\{\}$ as well as $I'\setminus I\ne\{\}$ then there exists no rate vector $\Rol'\in \Bcal^*$ such that (\ref{eq:I}) and (\ref{eq:I'}) hold simultaneously. To prove this, first we assume that (\ref{eq:I}) and (\ref{eq:I'}) hold for some $\Rol'\in \Bcal^*$ and some $(I,I')$ with the above property and then detect a contradiction. 

Two cases arise
depending on whether $I$ and $I'$ are disjoint or not. 
First consider the case where $I\cap I'=\{\}$. Using (\ref{eq:rule'}) in (\ref{eq:rule1}), we obtain
\bea
\nonumber
I\left(\Yol_{I\cup I'};\Zol_{I\cup I'}|\Zol_{{(I\cup I')}^c},V\right)
&=& I\left(\Yol_{I};\Zol_{I}|\Zol_{{I}^c},V\right)+
I\left(\Yol_{I'};\Zol_{I'}|\Zol_{{I'}^c},V\right)\\
\label{eq:suru1}
&&
\qquad +I\left(\Zol_I;\Zol_{I'}|\Zol_{{(I\cup I')}^c},V\right).
\eea
Now, adding (\ref{eq:I}) and (\ref{eq:I'}) and comparing with (\ref{eq:suru1}), we have
\bea
\nonumber
R'_{I\cup I'} &=& 
I\left(\Yol_{I\cup I'};\Zol_{I\cup I'}|\Zol_{{(I\cup I')}^c},V\right)
- I\left(\Zol_I;\Zol_{I'}|\Zol_{{(I\cup I')}^c},V\right)\\
\label{eq:suru2}
&<& I\left(\Yol_{I\cup I'};\Zol_{I\cup I'}|\Zol_{{(I\cup I')}^c},V\right) 
\eea
because $(\Zol_I,(\Zol_{{(I\cup I')}^c},V), \Zol_{I'})$ does not form  Markov chain. 
Note that (\ref{eq:suru2}) contradicts (\ref{eq:Rdef}) where of course $I\cup I'$ now plays the role of $I$.

Next consider the case where $I\cap I' = \Itil \ne \{\}$. Writing $I=(I\setminus \Itil)\cup \Itil$, from (\ref{eq:I}), we have 
\bea
\nonumber
R'_{I\setminus \Itil} + R'_{\Itil} &=& I\left(\Yol_{I};\Zol_I|\Zol_{I^c},V\right)\\
\label{eq:suru3}
&=& 
I\left(\Yol_{I\setminus \Itil};\Zol_{I\setminus \Itil}|\Zol_{{(I\setminus \Itil)}^c},V\right) + I\left(\Yol_{\Itil};\Zol_{\Itil}|\Zol_{\Itil^c},V\right)
+I\left(\Zol_{I\setminus \Itil};\Zol_{\Itil}|\Zol_{I^c},V\right)\qquad
\eea
which is same as (\ref{eq:suru1}) with $(I\setminus\Itil,\Itil)$ in place of $(I,I')$. Further, from (\ref{eq:Rdef}), note that 
\be
\label{eq:suru4}
R'_{\Itil} \ge I\left(\Yol_{\Itil};\Zol_{\Itil}|\Zol_{\Itil^c},V\right).
\ee
Using (\ref{eq:suru4}) in (\ref{eq:suru3}), we have
\be
\label{eq:suru5}
R'_{I\setminus \Itil} 
\le 
I\left(\Yol_{I\setminus \Itil};\Zol_{I\setminus \Itil}|\Zol_{{(I\setminus \Itil)}^c},V\right) + I\left(\Zol_{I\setminus \Itil};\Zol_{\Itil}|\Zol_{I^c},V\right).
\ee
Adding (\ref{eq:suru5}) and (\ref{eq:I'}) and comparing with (\ref{eq:suru1}) (with $I\setminus\Itil$ now in place of $I$), we obtain
\bea
\nonumber
R'_{I\cup I'} &\le& 
I\left(\Yol_{I\cup I'};\Zol_{I\cup I'}|\Zol_{{(I\cup I')}^c},V\right)\\
\label{eq:suru6}
&&\qquad
- I\left(\Zol_{I\setminus \Itil};\Zol_{I'}|\Zol_{{(I\cup I')}^c},V\right) 
+
I\left(\Zol_{I\setminus \Itil};\Zol_{\Itil}|\Zol_{I^c},V\right). 
\eea
Further, expand
\bea
\label{eq:suru7}
I\left(\Zol_{I\setminus \Itil};\Zol_{I'}|\Zol_{{(I\cup I')}^c},V\right)
&=& H\left(\Zol_{I\setminus \Itil}|\Zol_{{(I\cup I')}^c},V\right)
- H\left(\Zol_{I\setminus \Itil}|\Zol_{I'\cup{(I\cup I')}^c},V\right)\\
\label{eq:suru8} 
I\left(\Zol_{I\setminus \Itil};\Zol_{\Itil}|\Zol_{I^c},V\right)
&=&
H\left(\Zol_{I\setminus \Itil}|\Zol_{I^c},V\right)
- H\left(\Zol_{I\setminus \Itil}|\Zol_{\Itil\cup I^c},V\right),
\eea
note $I'\cup{(I\cup I')}^c = \Itil\cup I^c$ and subtract (\ref{eq:suru8}) from (\ref{eq:suru7}) to obtain
\bea
\nonumber
I\left(\Zol_{I\setminus \Itil};\Zol_{I'}|\Zol_{{(I\cup I')}^c},V\right)
- I\left(\Zol_{I\setminus \Itil};\Zol_{\Itil}|\Zol_{I^c},V\right)
&=& H\left(\Zol_{I\setminus \Itil}|\Zol_{{(I\cup I')}^c},V\right)
- H\left(\Zol_{I\setminus \Itil}|\Zol_{I^c},V\right)\\
\label{eq:suru9}
&=& I\left(\Zol_{I\setminus \Itil};\Zol_{I'\setminus \Itil}|\Zol_{{(I\cup I')}^c},V\right)\\
\label{eq:suru10}
&>& 0.
\eea
Here (\ref{eq:suru9}) follows by noting $I^c= {(I\cup I')}^c \cup (I'\setminus \Itil)$ and (\ref{eq:suru10}) follows due to the fact that $(\Zol_{I\setminus \Itil}, (\Zol_{{(I\cup I')}^c},V), \Zol_{I'\setminus \Itil})$ does not form Markov chain. Using (\ref{eq:suru10}) in (\ref{eq:suru6}), we obtain (\ref{eq:suru2}) again which contradicts (\ref{eq:Rdef}) as earlier. Hence the result. \hfill $\Box$

\subsection{Identifying Corner Points} 
\label{sec:number}

\begin{lemma}
\label{le:num} $\Bcal^*$ has at most $M'!$ corner points.
\end{lemma}

{\bf {\em Proof}:} Due to Lemma \ref{le:unique}, the number of corner points of $\Bcal^*$ is upper bounded by the number of possible ways we can have
$$ I^{(1)}\subset I^{(2)} \subset ... \subset I^{(m)} \subset I^{(m+1)}
\subset ... \subset I^{(M'-1)}\subset I_{M'},$$ $|I^{(m)}|=m$, $1\le m < M'$. Note that, given $I^{(m+1)}$, we can choose $I^{(m)}$ in $m+1$ possible ways. Consequently, we can choose $\{I^{(m)}\}_{m=1}^{M'-1}$ in $2\times 3\times ...\times M' = M'!$ possible ways. Hence the result. \hfill$\Box$

\begin{lemma}
\label{le:cornQ}
The rate vector $\Rol'$ such that
\be
\label{eq:cornQ}
R'_{i} = I(Y_{i};Z_{i}|\Zol_{I_{i-1}},V), \quad i\in I_{M'}
\ee
gives a corner point of $\Bcal^*$.
\end{lemma}

{\bf {\em Proof}:} From (\ref{eq:cornQ}), we can write
\be
\label{eq:cornQ1}
\sum_{i=m+1}^{M'} I(Y_{i};Z_{i}|\Zol_{I_{i-1}},V) = \sum_{i=m+1}^{M'} R'_i
\ee
for each $m=0,1,...,M'-1$.
Further, by Corollary \ref{cor:chain3}, (\ref{eq:cornQ1}) is same as
\be
\label{eq:cc3'}
I\left(\Yol_{I_{M'}\setminus I_m};\Zol_{I_{M'}\setminus I_m}|
\Zol_{I_{m}},V\right)= R'_{I_{M'}\setminus I_m}
\ee   
which makes $M'$ constraints, given in (\ref{eq:Rdef}), active. To see this, set $I= I_{M'}\setminus I_m$ in (\ref{eq:Rdef}), vary $m=0,1,...,M'-1$ and compare with (\ref{eq:cc3'}). 
Therefore, in order to establish $\Rol'$ as a corner point of $\Bcal^*$, we are just left to show $\Rol'\in \Bcal^*$. Towards that, note, by Lemma \ref{le:last} and (\ref{eq:cornQ}), that  
\be
\label{eq:lastL}
I\left(\Yol_{I};\Zol_{I}|\Zol_{I^c},V\right) \le
\sum_{i\in I} I\left(Y_i;Z_i|\Zol_{I_{i-1}},V\right) = R'_I
\ee
for all $I\subseteq I_{M'}\setminus\{\}$. Comparing (\ref{eq:lastL}) with (\ref{eq:Rdef}), we conclude $\Rol'\in \Bcal^*$. This completes the proof. \hfill$\Box$
 
{\bf {\em Proof of Lemma \ref{le:corner}}:} Now, noting symmetry, the indices $\{1,2,...,M'\}$ in (\ref{eq:cornQ}) can be permuted to obtain $M'!$ corner points. Finally, by Lemma \ref{le:num}, such enumeration of corner points of $\Bcal^*$ is exhaustive. Hence the result. \hfill $\Box$

\end{document}